\documentclass{emulateapj}

\slugcomment{Submitted to ApJ 16 July 2006, Accepted 3 October 2006, Revised 2 November 2006}
\shorttitle{Transience of hot dust around sun-like stars}
\shortauthors{Wyatt et al.}

\begin{document}

\title{Transience of hot dust around sun-like stars}

\author{M. C. Wyatt}
\affil{Institute of Astronomy, University of Cambridge,
  Madingley Road, Cambridge CB3 0HA, UK}
\email{wyatt@ast.cam.ac.uk}

\author{R. Smith}
\affil{Institute for Astronomy, Royal Observatory,
  Blackford Hill, Edinburgh EH9 3HJ, UK}

\author{J. S. Greaves}
\affil{Scottish Universities Physics Alliance, University of St. Andrews, 
  Physics \& Astronomy, North Haugh, St Andrews KY16 9SS, UK}

\author{C. A. Beichman\altaffilmark{1} and G. Bryden}
\affil{Jet Propulsion Laboratory, 4800 Oak Grove Drive, Pasadena, CA 91109, USA}

\author{C. M. Lisse}
\affil{Planetary Exploration Group, Space Department, Johns Hopkins University
  Applied Physics Laboratory, 11100 Johns Hopkins Rd, Laurel, MD 20723, USA}

\altaffiltext{1}{Michelson Science Center, California Institute of Technology, M/S 100-22,
  Pasadena, CA 91125, USA}

\begin{abstract}
There is currently debate over whether the dust content of planetary systems
is stochastically regenerated or originates in planetesimal belts evolving in
quasi-steady state.
In this paper a simple model for the steady state evolution of debris
disks due to collisions is developed and confronted with the properties of
the emerging population of 7 sun-like stars that have hot dust at $<10$ AU.
The model shows that there is a maximum possible disk mass at a given age,
since more massive primordial disks process their mass faster.
The corresponding maximum dust luminosity is $f_{\rm{max}}=0.16\times 
10^{-3}r^{7/3}t_{\rm{age}}^{-1}$, where $r$ is disk radius in AU and $t_{\rm{age}}$ is
system age in Myr.
The majority (4/7) of the hot disks exceed this limit by a factor $\gg 1000$ and so
cannot be the products of massive asteroid belts, rather the following systems must be
undergoing transient events characterized by an unusually high dust content near
the star:
$\eta$ Corvi, HD69830, HD72905 and BD+20307.
It is also shown that the hot dust cannot originate in a recent collision in an
asteroid belt, since there is also a maximum rate at which collisions of sufficient magnitude
to reproduce a given dust luminosity can occur in a disk of a given age.
For the 4 transient disks, there is at best a 1:$10^5$ chance of witnessing
such an event compared with 2\% of stars showing this phenomenon.
Further it is shown that the planetesimal belt feeding the dust in these
systems must be located further from the star than the dust, typically at $\gg 2$ AU.
Other notable properties of the 4 hot dust systems are:
two also have a planetesimal belt at $>10$ AU ($\eta$ Corvi and HD72905);
one has 3 Neptune mass planets at $<1$ AU (HD69830);
all exhibit strong silicate features in the mid-IR.
We consider the most likely origin for the dust in these systems to be a dynamical
instability which scattered planetesimals inwards from a more distant planetesimal belt
in an event akin to the Late Heavy Bombardment in our own system, the dust being
released from such planetesimals in collisions and possibly also sublimation.
Further detailed study of the planet, planetesimal and dust populations in these
rare objects has the potential to uncover the chaotic evolutionary history of these
systems and to shed light on the history of the solar system.
\end{abstract}

\maketitle

\section{Introduction}
\label{s:intro}
Planetesimal belts appear to be a common feature of planetary systems.
There are two main belts in the solar system: the asteroid belt and
the Kuiper belt.
These belts inhabit the regions of the solar system where planetesimal
orbits can remain stable over the 4.5 Gyr age of our system (Lecar et al. 2001).
The larger planetesimals in the belts are continually grinding
down feeding the smaller bodies in a process known as a collisional cascade
which is slowly eroding the belts (Bottke et al. 2005).
The smallest dust in the asteroid belt is acted on by radiation forces;
P-R drag makes the dust spiral in toward the Sun making a disk known as
the zodiacal cloud that the Earth sits in the middle of (Leinert \& Gr\"{u}n 1990).
A dust cloud is also predicted to arise from collisions amongst Kuiper belt
objects (Liou \& Zook 1999), although our information
on this population is sparse (Landgraf et al. 2002)
because its emission is masked by the zodiacal emission
(Backman, Dasgupta \& Stencel 1995) and few dust grains make it into the
inner solar system (Moro-Mart\'{\i}n \& Malhotra 2003).

Many extrasolar systems also have such planetesimal belts, known as
debris disks.
These have been detected from their dust content (Aumann et al. 1984) from which
it has been inferred that larger planetesimals must exist to replenish
the dust disks because of the short lifetime of this dust (Backman \& Paresce 1993).
The collisional cascade scenario is supported by modeling of the
emission spectrum of the dust which shows a size distribution similar
to that expected for dust coming from a collisional cascade (Wyatt \& Dent 2002, hereafter
WD02).
However, the issue of how these disks evolve has recently come under
close scrutiny.

From a theoretical point view, Dominik \& Decin (2003; hereafter DD03)
showed that if P-R drag is not important then a planetesimal belt
evolving in quasi-steady state would lose mass due to collisional grinding
down giving a disk mass (and dust luminosity) that falls off $\propto t^{-1}$.
This is in broad agreement with the observed properties of debris disks:
the mean dust luminosity at a given age falls off $\propto t^{-1.8}$
(Spangler et al. 2001);
the mass inferred from detection statistics falls off $\propto t^{-0.5}$
(Greaves \& Wyatt 2003), while the mass of the detected disks falls off
$\propto t^{-1}$ (Najita \& Williams 2005);
the upper limit in luminosity of the detected disks also falls off $\propto t^{-1}$
(Rieke et al. 2005).
While these trends can be viewed as a success of the steady-state model,
it has yet to be proved that a steady state evolution model fits the data
in more than just general terms (Meyer et al. 2006).
Several puzzling observations also remain to be explained.

Decin et al. (2003) noted that the maximum fractional luminosity of debris disks remains
constant at $f=L_{\rm{ir}}/L_\star \approx 10^{-3}$ up to the oldest stars,
where $L_{\rm{ir}}$ and $L_\star$ are the disk and stellar luminosities respectively
(see also Table \ref{tab:symb} for definitions of the parameters used in the text), and this
was explained by DD03 as a consequence of delayed stirring.
A delay in the ignition of a collisional cascade is expected if 
it is the formation of Pluto-sized objects which trigger the cascade, since
such massive bodies take longer, up to several Gyr, to form
further from the star (Kenyon \& Bromley 2002).
However, that interpretation predicts that the radius of the belts should increase with stellar 
age, and this is not observed (Najita \& Williams 2005).
There is also recent evidence that the dust content of some systems is transient.
The discovery of a population of dust grains around Vega in the process of removal by 
radiation pressure indicates that this system cannot have remained in steady state for the full 
350 Myr age of the star (Su et al. 2005).
Rieke et al. (2005) used their statistics on A stars, which showed a wide variety of properties 
among the debris disks, to suggest that much of the dust we see is produced episodically in 
collisions between large planetesimals.
There is also an emerging population of debris disks detected around sun-like
stars with dust at a few AU (Gaidos 1999; Beichman et al. 2005; Song et al. 2005;
Smith, Wyatt \& Dent in prep.).
There is debate over whether these are atypically massive asteroid belts or the
consequence of a rare transient event (e.g., Beichman et al. 2005).

A stochastic element to the evolution of debris disks would fit with 
our understanding of the evolution of the dust content of the inner 
solar system.
This is believed to have been significantly enhanced for timescales of a few
Myr following collisions between objects $\sim 100$ km in size in the asteroid 
belt (Nesvorn\'{y} et al. 2003; Farley et al. 2006).
However, it is not known whether the aftermath of individual collisions would be
detectable in a debris disk, or indeed whether such events would happen frequently
enough to explain the statistics (WD02; Telesco et al. 2005).
Such events have a dramatic effect on the amount of dust in the solar system
because there is relatively little around during the quiescent periods.
Planetesimal belts of equivalent mass to those in the solar system would
not have been detected in the current debris disk surveys.
However, there is evidence to suggest that both belts were $\sim 200$ times more
massive in the past (e.g., Stern 1996; Bottke et al. 2005).
Periods analogous to the heavy bombardment experienced in the solar system up to
$\sim 700$ Myr after its formation have also been invoked to explain the fact that debris
disks are most often detected around stars $<400$ Myr old (Habing et al. 1999).

In the light of this controversy we revisit a simple analytical model
for the steady state collisional evolution of planetesimal belts which was
originally explored in DD03.
The model we derive for that evolution is given in \S \ref{s:model},
and differs in a subtle but important way from that of DD03, since it affects
the dust production as a function of collision velocity.
This model shows that there is a maximum possible
disk mass (and dust luminosity) at any given age.
In \S \ref{s:hot} confrontation with the few hot planetesimal belts discovered recently
shows that the majority of these cannot be explained as massive asteroid belts,
rather these must be systems undergoing a transient event.
The possibility that these are caused by a recent collision within a planetesimal
belt is also discussed, as is the possibility that the dust originates in a planetesimal
belt in the terrestrial planet region.
The implications of these results are discussed in \S \ref{s:conc}.
Application of the model to the statistics of detected debris disks will be
considered in a later paper (Wyatt et al., in prep.).

\section{Analytical collisional evolution model}
\label{s:model}
In this section a simple analytical model is developed for the
evolution of a planetesimal belt due to collisions amongst
its members.
The parameters used in this model are summarized in the table \ref{tab:symb}
which also gives the units assumed for these parameters throughout the paper.

\subsection{The planetesimal belt size distribution}
\label{ss:pb}
The planetesimal belt is assumed to be in collisional
equilibrium with a size distribution defined by:
\begin{equation}
  n(D) = K D^{2-3q}, \label{eq:nd}
\end{equation}
where $q=11/6$ in an infinite collisional cascade (Dohnanyi 1969)
and the scaling parameter $K$ is called $f_a$ by DD03.
That distribution is assumed to hold from the largest planetesimal
in the disk, of diameter $D_{\rm{c}}$, down to the size below which particles
are blown out by radiation pressure as soon as they are created,
$D_{\rm{bl}}$.
If we assume that $q$ is in the range 5/3 to 2 then most of the
mass is in the largest planetesimals while the cross-sectional area is
in the smallest particles such that:
\begin{eqnarray}
  \sigma_{\rm{tot}} & = & 3.5 \times 10^{-17} K(3q-5)^{-1} (10^{-9}D_{\rm{bl}})^{5-3q} \label{eq:stot} \\
  M_{\rm{tot}}     & = & 8.8 \times 10^{-17} K \rho (6-3q)^{-1} D_{\rm{c}}^{6-3q}, \label{eq:mtot1} \\
               & = & 2.5 \times 10^{-9} \left( \frac{3q-5}{6-3q} \right)
                   \rho \sigma_{\rm{tot}}D_{\rm{bl}}
                  \left( \frac{10^9 D_{\rm{c}}}{D_{\rm{bl}}} \right)^{6-3q}, 
     \label{eq:mtot2}
\end{eqnarray}
where spherical particles of density $\rho$ have been assumed and $M_{\rm{tot}}$ is
in $M_\oplus$ if the units of table \ref{tab:symb} are used for the other parameters.

The planetesimal belt is assumed to be at a radius $r$, and to have a width $dr$ (in AU).
One of the observable properties of a planetesimal belt is its fractional luminosity,
$f=L_{\rm{ir}}/L_\star$, i.e., the infrared luminosity from the disk divided by the stellar
luminosity.
Assuming that the grains act like black bodies and so absorb all the radiation
they intercept we can write:
\begin{equation}
  f = \sigma_{\rm{tot}}/(4\pi r^2). \label{eq:f}
\end{equation}
In other words, in this model $\sigma_{\rm{tot}}$, $M_{\rm{tot}}$ and $f$ 
are all proportional to each other and just one is needed to define the scaling
factor $K$ in equation (\ref{eq:nd}).
Assuming the particles act like black bodies also allows us to derive
the following relation:
\begin{equation}
  D_{\rm{bl}} = 0.8(L_\star/M_\star)(2700/\rho), \label{eq:dbl}
\end{equation}
where $D_{\rm{bl}}$ is in $\mu$m, $L_\star$ and $M_\star$ are in solar units, and
$\rho$ is in kg m$^{-3}$.

Relaxing the black body assumption is easily achieved (e.g., WD02).
However, this would result in relatively small changes in the way $f$ 
scales with $M_{\rm{tot}}$, and so for its heuristic simplicity we keep this assumption
throughout this paper.
Probably the most important simplification within this model is that of the
continuous size distribution.
For example, we know that the cut-off in the size distribution at $D_{\rm{bl}}$ would
cause a wave in the size distribution at sizes just larger than this (Th\'{e}bault,
Augereau \& Beust 2003), that large quantities of blow-out grains can also affect
the distribution of small size particles (Krivov, Mann, \& Krivova 2000), and that
the dependence of planetesimal strength on size can result in $q \ne 11/6$ as well
as a wave in the distribution at large sizes (Durda et al. 1998;
O'Brien \& Greenberg 2003).
Also, since the largest planetesimals would not be in collisional equilibrium
at the start of the evolution, their initial distribution may not be the same
as that of a collisional cascade, although distributions with $q \approx 11/6$
have been reported from planet formation models (e.g., Stern \& Colwell 1997;
Davis \& Farinella 1997; Kenyon \& Luu 1999) meaning this is a reasonable starting
assumption.
Despite these simplifications, we believe this model is adequate to explore to 
first order the evolution of planetesimal belts which can later be studied in
more depth.

\subsection{Collisional evolution}
\label{ss:ce}
In a collisional cascade material in a bin with a given size range $D$ to $D+dD$ is replaced
by fragments from the destruction of larger objects at the same rate that it is destroyed
in collisions with other members of the cascade.
The long-timescale evolution is thus determined by the removal of mass from the top end of
the cascade.
In this model the scaling factor $K$ (and so the total mass and fractional luminosity
etc) decreases as the number of planetesimals of size $D_{\rm{c}}$ decreases.
The loss rate of such planetesimals is determined by their collisional lifetime, which in
the terminology of WD02 is given by:
\begin{equation}
  t_{\rm{c}} = \sqrt{r^3/M_\star} (r dr/\sigma_{\rm{tot}}) [2I/f(e,I)] / f_{\rm{cc}}, 
\label{eq:tc1}
\end{equation}
where maintaining the units used previously gives $t_{\rm{c}}$ in years, $I$ is the mean 
inclination
of the particles' orbits (which determines the torus height), $f(e,I)$ is the ratio of
the relative velocity of collisions to the Keplerian velocity ($=v_{\rm{rel}}/v_{\rm{k}}$, 
also called $\nu$ by DD03),
and $f_{\rm{cc}}$ is the fraction of the total cross-sectional area in the belt which
is seen by planetesimals of size $D_{\rm{c}}$ as potentially causing a catastrophic 
collision.

From hereon we will use the assumption that $f(e,I) = \sqrt{1.25e^2+I^2}$, where $e$
is the mean eccentricity of the particles, which is valid for Rayleigh distributions
of $e$ and $I$ (Lissauer \& Stewart 1993; Wetherill \& Stewart 1993).
An expression for $f_{\rm{cc}}$ was given in WD02, however, here we will
ignore the gravitational focussing effect, which is important in the
accumulation phase but not during the destruction
phase of a planetesimal belt (see \S \ref{ss:pc}),
and so derive an expression that is the
same as that given in Wyatt et al. (1999):
\begin{equation}
  f_{\rm{cc}} = (10^{-9} D_{\rm{bl}}/D_{\rm{c}})^{3q-5}G(q,X_{\rm{c}}), \label{eq:fcc}
\end{equation}
where $X_{\rm{c}}=D_{\rm{cc}}/D_{\rm{c}}$, $D_{\rm{cc}}$ is the smallest planetesimal 
that has enough energy to catastrophically destroy a planetesimal of size $D_{\rm{c}}$ (which 
is
called $\epsilon$ in DD03), and:
\begin{eqnarray}
  G(q,X_{\rm{c}}) & = & [(X_{\rm{c}}^{5-3q}-1)+ (6q-10)(3q-4)^{-1}(X_{\rm{c}}^{4-3q}-1) 
\nonumber \\
           &   & + (3q-5)(3q-3)^{-1}(X_{\rm{c}}^{3-3q}-1)]. \label{eq:qgxc}
\end{eqnarray}

The factor $X_{\rm{c}}$ can be worked out from the dispersal threshold, $Q_{\rm{D}}^\star$, 
defined
as the specific incident energy required to catastrophically destroy a particle such 
that (WD02):
\begin{eqnarray}
  X_{\rm{c}} & = & (2Q_{\rm{D}}^\star/v_{\rm{rel}}^2)^{1/3}, \label{eq:xc1} \\
      & = & 1.3 \times 10^{-3} [Q_{\rm{D}}^\star r M_\star^{-1} f(e,I)^{-2} ]^{1/3}, 
\label{eq:xc2}
\end{eqnarray}
where $Q_{\rm{D}}^\star$ is in J kg$^{-1}$ (called $S$ in DD03\footnote{Equation 25 in DD03 differs 
from our
equation (\ref{eq:xc1}) because we define $Q_{\rm{D}}^\star$ to be the specific incident 
kinetic
energy so that $0.5M_2v_{\rm{rel}}^2=M_1 Q_{\rm{D}}^\star$ whereas DD03 define $S$ to 
be the specific binding energy of the two objects (giving their equation 24).
In the limit of $S \ll v_{\rm{rel}}^2/8$ the two equations are the same, since $X_{\rm{c}} 
\ll 
1$.})

Combining the above equations gives for the collisional lifetime of the planetesimals
of size $D_{\rm{c}}$:
\begin{eqnarray}
  t_{\rm{c}} & = & \left( \frac{r^{2.5} dr}{M_\star^{0.5} \sigma_{\rm{tot}}} \right) 
            \left( \frac{2[1+1.25(e/I)^2]^{-0.5}}{G(q,X_{\rm{c}})} \right)
            \left( \frac{10^{-9}D_{\rm{bl}}}{D_{\rm{c}}} \right)^{5-3q}, 
\label{eq:tcstot} \\
      & = & \left( \frac{3.8\rho r^{2.5} dr 
            D_{\rm{c}}}{M_\star^{0.5} M_{\rm{tot}}}  \right) 
            \left( \frac{(12q-20)[1+1.25(e/I)^2]^{-0.5}}{(18-9q)G(q,X_{\rm{c}})} \right).
            \label{eq:tcmtot}
\end{eqnarray}
Assuming that collisions are the only cause of mass loss in the belt, the evolution of
the disk mass $M_{\rm{tot}}(t)$ (or equivalently of $K$, $\sigma_{\rm{tot}}$, or $f$)
can be worked out by solving $dM_{\rm{tot}}/dt = -M_{\rm{tot}}/t_{\rm{c}}$ to give:
\begin{equation}
  M_{\rm{tot}}(t) = M_{\rm{tot}}(0)/[1+t/t_{\rm{c}}(0)], \label{eq:mtott}
\end{equation}
where $M_{\rm{tot}}(0)$ is the initial disk mass and $t_{\rm{c}}(0)$ is the collisional 
lifetime at that initial epoch;
this solution is valid as long as mass is the only parameter of the planetesimal belt
that changes with time.
This results in a disk mass which is constant at $M_{\rm{tot}}(0)$ for $t \ll t_{\rm{c}}(0)$, 
but which falls off $\propto 1/t$ for $t \gg t_{\rm{c}}(0)$ (as noted, e.g., in DD03).

\begin{figure*}
  \centering
  \begin{tabular}{cc}
     \hspace{-0.15in} 
       \includegraphics[width=3.2in]{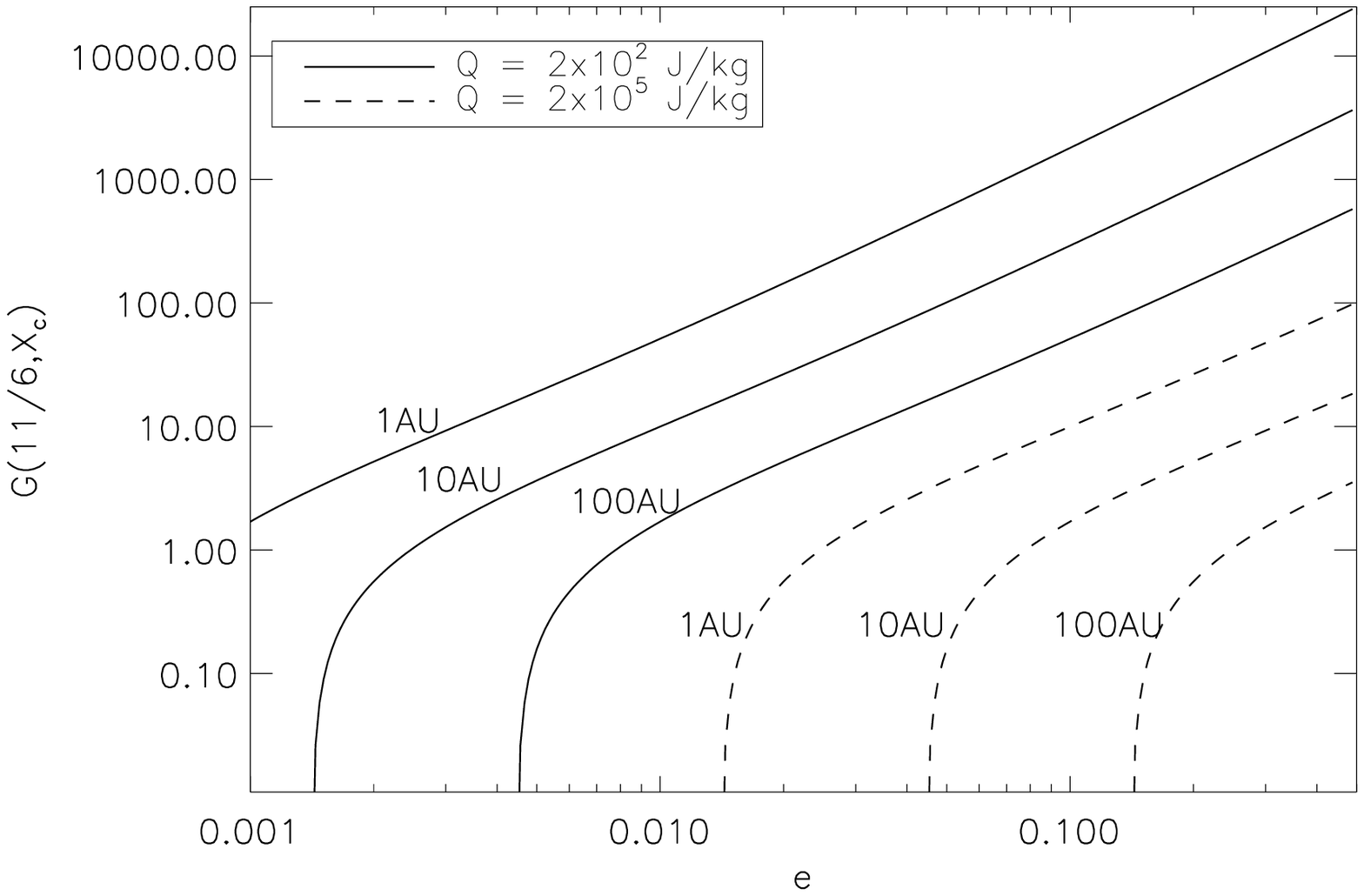} &
     \hspace{-0.15in} 
       \includegraphics[width=3.2in]{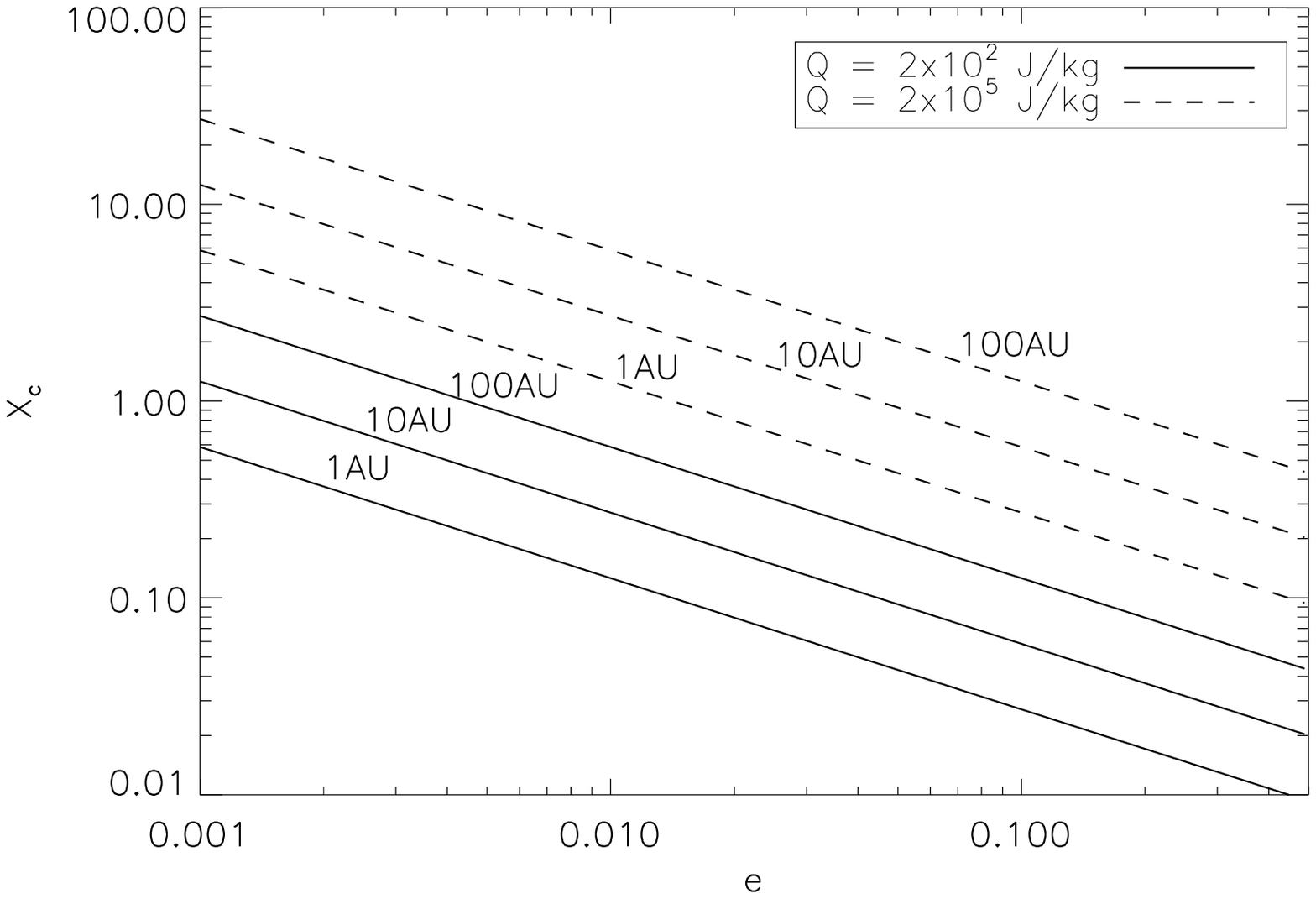}
  \end{tabular}
  \caption{The dependence of \textbf{(left)} $G(11/6,X_{\rm{c}})$ and \textbf{(right)}
  $X_{\rm{c}}$ on planetesimal eccentricity ($e$) for planetesimals of different
  strengths ($Q_{\rm{D}}^\star$) and at different distances from the star ($r$).}
\label{fig:gvse}
\end{figure*}

However, another interesting property of this evolution is that, since the
expression for $t_{\rm{c}}(0)$ includes a dependence on $M_{\rm{tot}}(0)$, the disk
mass at late times is independent of initial disk mass.
This is because more massive disks process their mass faster.
This means that for any given age, $t_{\rm{age}}$, there is a maximum disk mass 
$M_{\rm{max}}$
(and also infrared luminosity, $f_{\rm{max}}$) that can remain due to collisional processing:
\begin{eqnarray}
  M_{\rm{max}} & = & \left( \frac{3.8 \times 10^{-6} \rho r^{3.5} (dr/r) 
                     D_{\rm{c}}}{M_\star^{0.5}t_{\rm{age}}} \right) \times \nonumber \\
          & &  \left( \frac{(12q-20)[1+1.25(e/I)^2]^{-0.5}}{(18-9q)G(q,X_{\rm{c}})} \right), 
            \label{eq:mmax1} \\
  f_{\rm{max}} & = & \left( \frac{10^{-6} r^{1.5}(dr/r)}{4\pi M_\star^{0.5} t_{\rm{age}}} \right)
                     \left( \frac{10^{-9} D_{\rm{bl}}}{D_{\rm{c}}} \right)^{5-3q}
                     \times \nonumber \\ 
          & &  \left( \frac{2[1+1.25(e/I)^2]^{-0.5}}{G(q,X_{\rm{c}})} \right).
             \label{eq:fmax1} 
\end{eqnarray}
In this model, the present day disk mass (or luminosity) is expected to
be equal to this "maximum" disk mass (or luminosity) for disks in which the
largest planetesimals are in collisional equilibrium.
This corresponds to disks around stars that are older than the collisional
lifetime of those planetesimals given in equation (\ref{eq:tcmtot}).

For example, with the further assumptions that $q=11/6$, $e \approx I$, and
$\rho = 2700$ kg m$^{-3}$, we find:
\begin{eqnarray}
  M_{\rm{max}} & = & 0.009 r^{3.5} (dr/r)
    D_{\rm{c}} M_\star^{-0.5} t_{\rm{age}}^{-1}/G(11/6,X_{\rm{c}}), \label{eq:mmax2} \\
  f_{\rm{max}} & = & 0.004 r^{1.5} (dr/r)
    D_{\rm{c}}^{0.5} L_\star^{-0.5} t_{\rm{age}}^{-1}/G(11/6,X_{\rm{c}}), \label{eq:fmax2}
\end{eqnarray}
where $M_{\rm{max}}$ is in $M_\oplus$, $r$ in AU, $D_{\rm{c}}$ in km, $t_{\rm{age}}$ in Myr, 
and
$G(11/6,X_{\rm{c}})=X_{\rm{c}}^{-0.5}+0.67X_{\rm{c}}^{-1.5}+0.2X_{\rm{c}}^{-2.5}-1.87$,
with $X_{\rm{c}}=10^{-3}(rQ_{\rm{D}}^\star/e^2)^{1/3}$ ($Q_{\rm{D}}^\star$ is in J kg$^{-1}$).

Plots of $G(11/6,X_{\rm{c}})$ and $X_{\rm{c}}$ for typical planetesimal belts are
shown in Fig.~\ref{fig:gvse}.
However, for many disks the approximation that $X_{\rm{c}} \ll 1$ is valid, and
so $G(11/6,X_{\rm{c}}) \approx 0.2X_{\rm{c}}^{-2.5} =
6.3 \times 10^6 r^{-5/6}{Q_{\rm{D}}^\star}^{-5/6}e^{5/3}M_\star^{5/6}$, giving:
\begin{eqnarray}
  M_{\rm{max}} & = & 1.4 \times 10^{-9} r^{13/3} (dr/r)
    D_{\rm{c}} {Q_{\rm{D}}^\star}^{5/6} \times \nonumber \\
    & & e^{-5/3} M_\star^{-4/3} t_{\rm{age}}^{-1}, \label{eq:mmax3} \\
  f_{\rm{max}} & = & 0.58 \times 10^{-9} r^{7/3} (dr/r)
    D_{\rm{c}}^{0.5} {Q_{\rm{D}}^\star}^{5/6}\times \nonumber \\
    & & e^{-5/3} M_\star^{-5/6} L_\star^{-0.5} t_{\rm{age}}^{-1}.
  \label{eq:fmax3}
\end{eqnarray}

\subsection{Comparison with DD03}
\label{ss:dd03}
\begin{figure*}
  \centering
  \begin{tabular}{cc}
     \hspace{-0.15in} \includegraphics[width=3.2in]{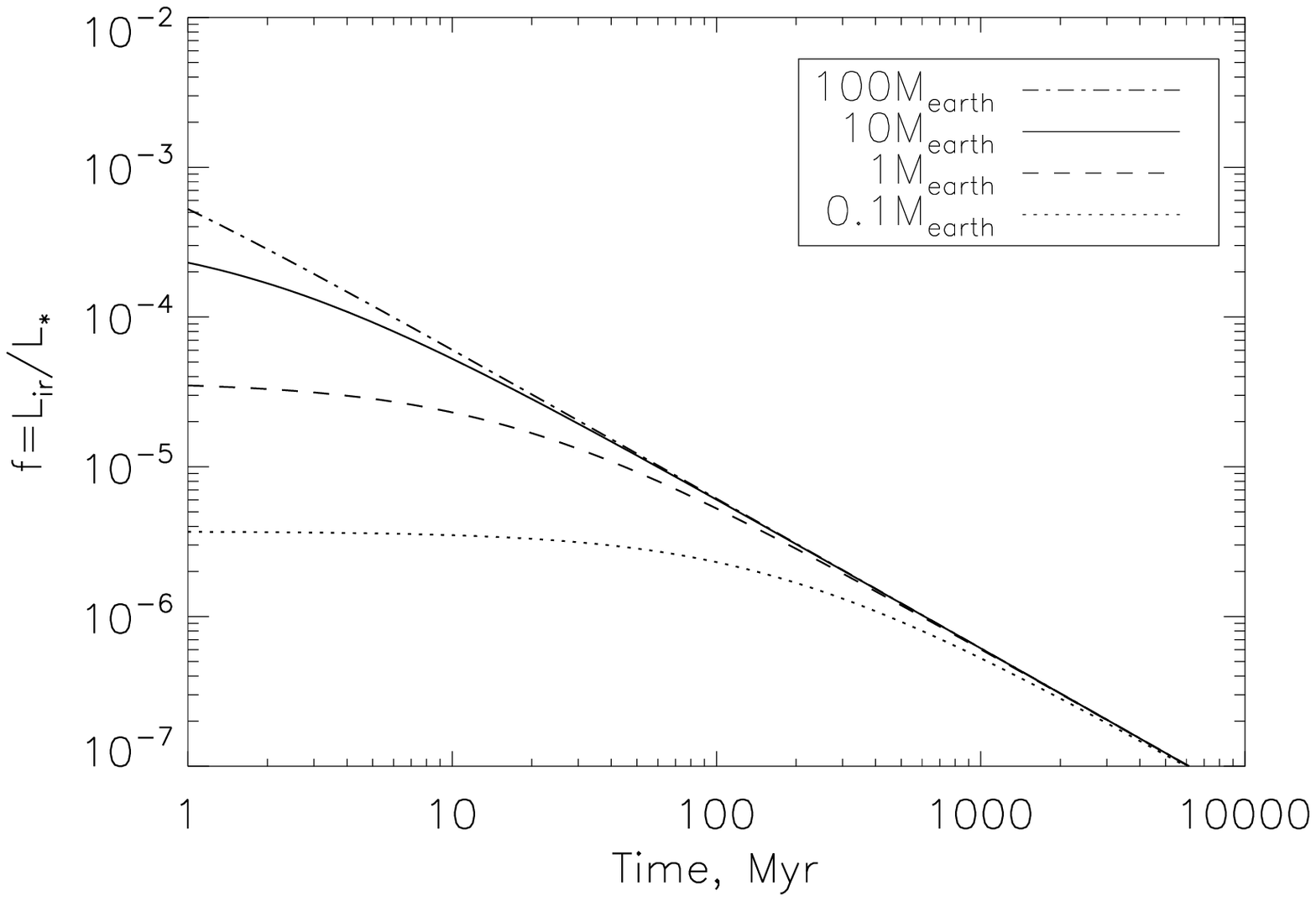} &
     \hspace{-0.15in} \includegraphics[width=3.2in]{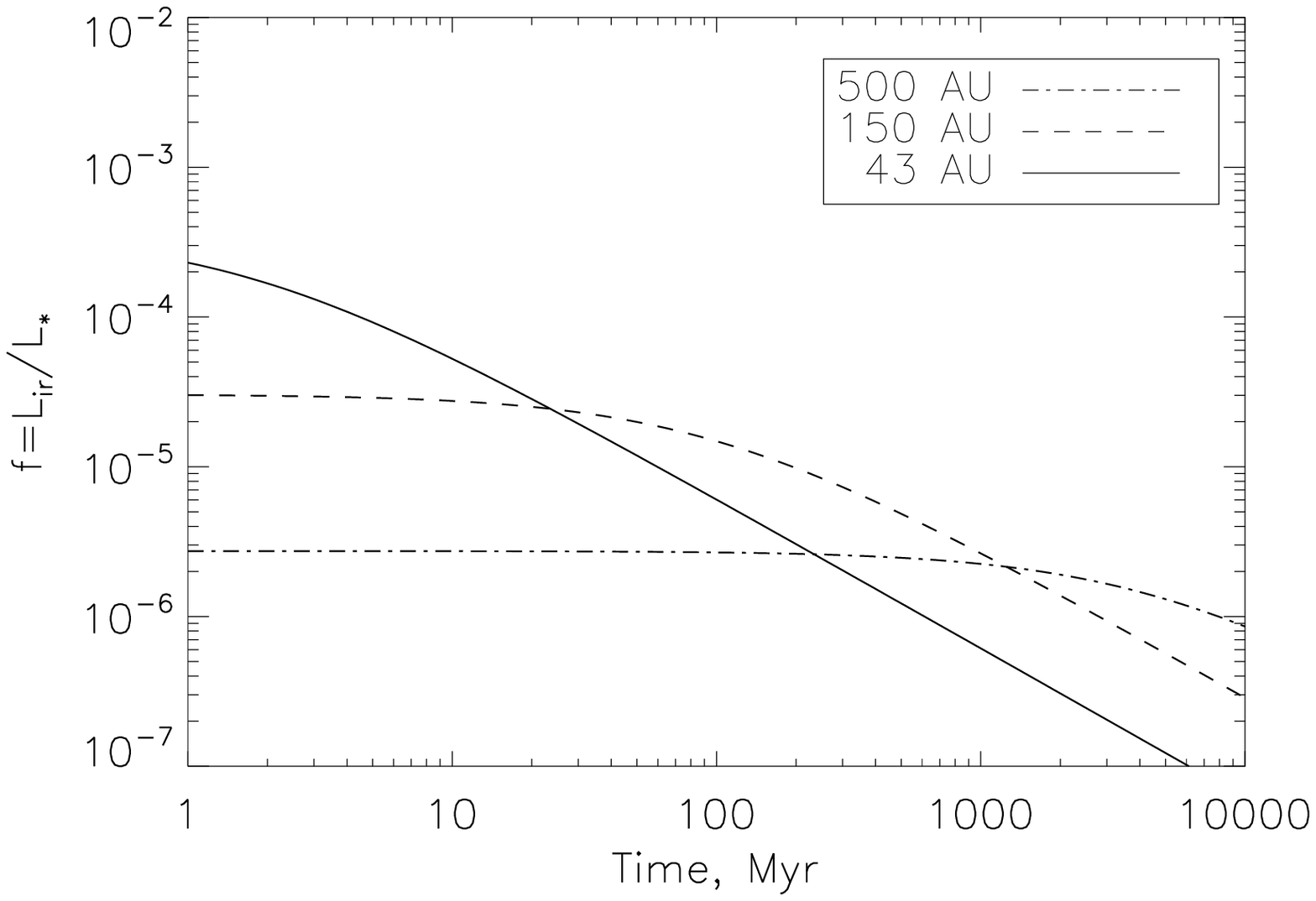} \\[-0.0in]
     \hspace{-0.15in} \includegraphics[width=3.2in]{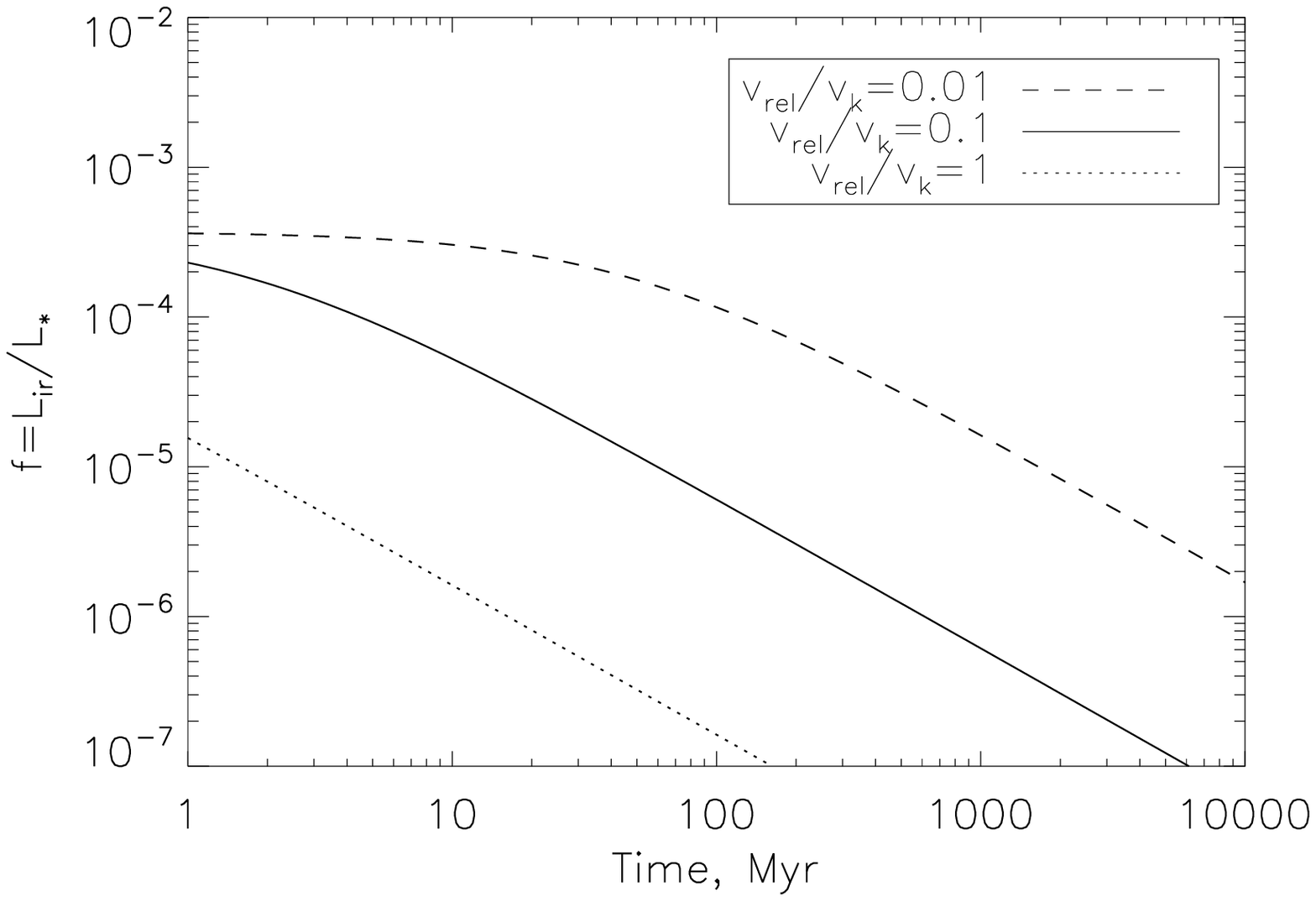} &
     \hspace{-0.15in} \includegraphics[width=3.2in]{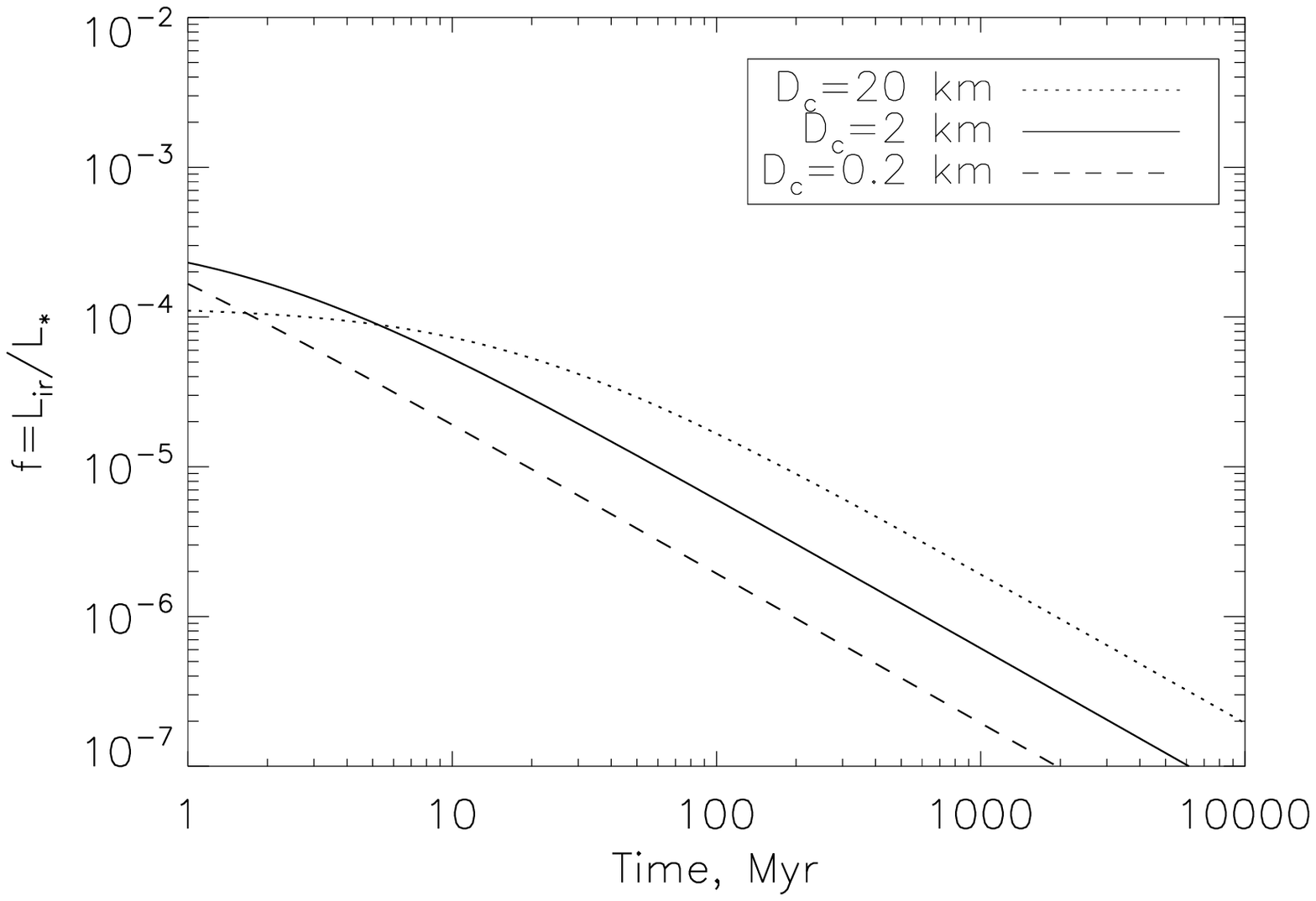} \\[-0.0in]
     \hspace{-0.15in} \includegraphics[width=3.2in]{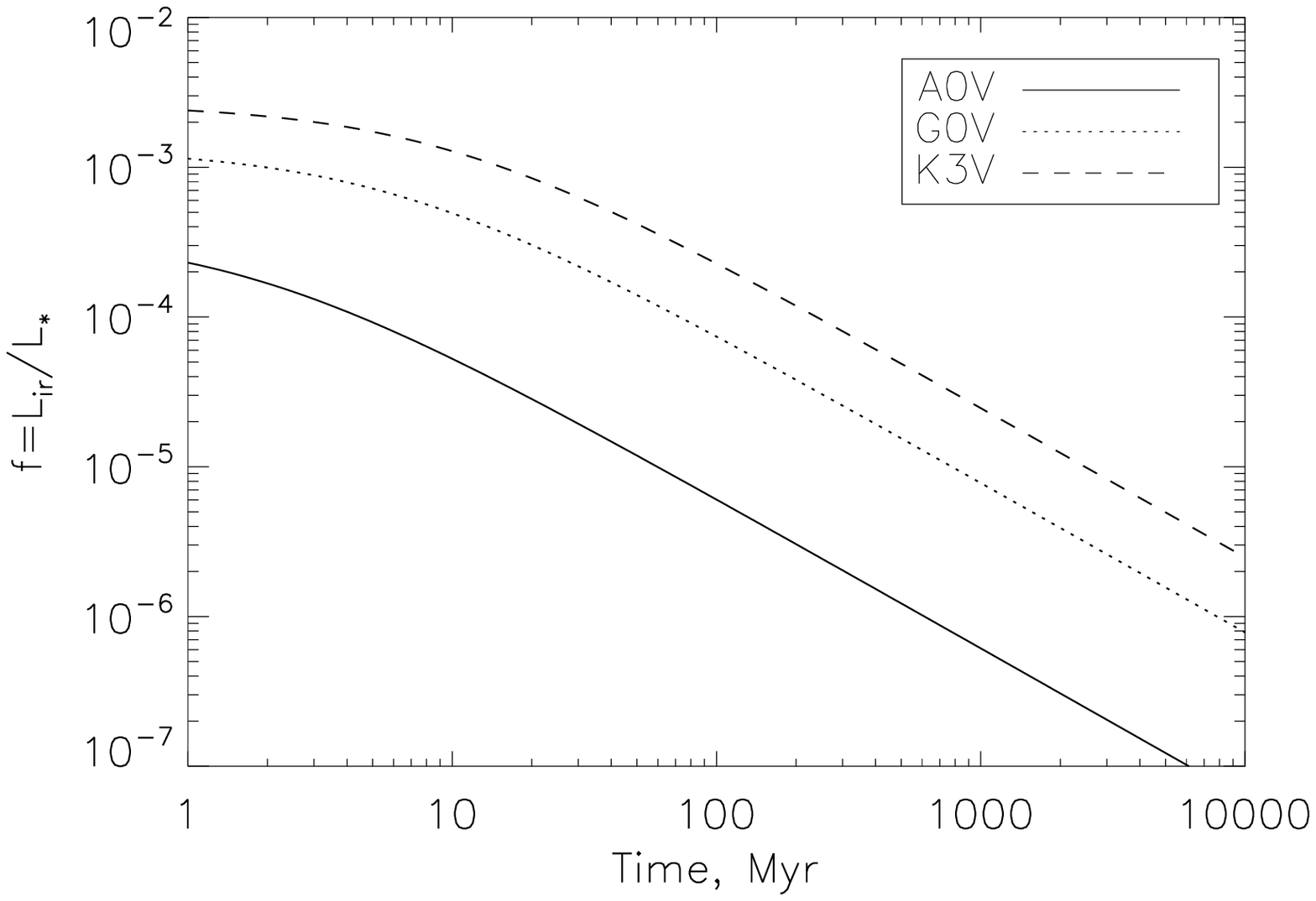} &
  \end{tabular}
  \caption{The collisional evolution of a planetesimal belt with parameters similar
  to the nominal model of DD03 
  [$r=43$ AU, $dr=15$ AU, $D_{\rm{c}}=2$ km, $\rho=2700$ kg m$^{-3}$, $f(e,I)=0.1$,
  $e/I=1$, $Q_{\rm{D}}^\star=200$ J kg$^{-1}$, $M_{\rm{tot}}(0)=10M_\oplus$, A0 star]
  showing the effect of changing:
  \textbf{(top left)} starting disk mass $M_{\rm{tot}}(0)$,
  \textbf{(top right)} disk radius $r$,
  \textbf{(middle left)} collision velocity $v_{\rm{rel}}/v_{\rm{k}}$,
  \textbf{(middle right)} maximum planetesimal size $D_{\rm{c}}$, and
  \textbf{(bottom left)} stellar spectral type.
  These plots can be directly compared to figs. 1b-f of DD03.}
  \label{fig:dd03}
\end{figure*}

Since DD03 produced a very similar analytical model, our results were
compared with those of DD03.
The results of disk evolution for a planetesimal belt close to their nominal
model were computed using the parameters:
$r=43$ AU, $dr=15$ AU, $D_{\rm{c}}=2$ km, $\rho=2700$ kg m$^{-3}$, $f(e,I)=0.1$, $e/I=1$, 
$Q_{\rm{D}}^\star=200$ J kg$^{-1}$, $M_{\rm{tot}}(0)=10M_\oplus$, A0 star (for which $L_\star=54L_\odot$,
$M_\star=2.9M_\odot$, $D_{\rm{bl}}=15$ $\mu$m).
Each of the parameters $M_{\rm{tot}}(0)$, $r$, $f(e,I)$, $D_{\rm{c}}$ and spectral type 
were also varied to make the plots shown in Fig.~\ref{fig:dd03} which are
equivalent to Figs 1b-1f of DD03.

The results are very similar in most regards:
more massive disks start out with higher $f$, but the turnover from constant to $1/t$
evolution is later for lower mass disks meaning that at late times all disks converge
to the same maximum value (Fig.~\ref{fig:dd03}a);
putting the same mass at larger distances reduces the initial dust luminosity $f$, but the
resulting lower surface density and longer orbital timescales there combine to make the
turnover happen later which means that at late times more distant belts are more 
massive (Fig.~\ref{fig:dd03}b);
putting the same mass into larger planetesimals reduces the cross-sectional area
of dust (equation \ref{eq:mtot2}) and so the initial dust luminosity $f$, but increases
the collisional lifetime of those planetesimals (equation \ref{eq:tcmtot}) which means
that at late times belts with larger planetesimals retain their mass for longer
(Fig.~\ref{fig:dd03}d);
later spectral types have higher starting dust luminosities because the cascade extends
down to smaller sizes (equation \ref{eq:dbl}), and the longer orbital times mean that
they keep their mass for longer (Fig.~\ref{fig:dd03}e).

Where the models differ is in the exact way $M_{\rm{tot}}$ is used to get $f$ and 
$t_{\rm{c}}$, and
in the way the evolution is affected by changing $v_{\rm{rel}}/v_{\rm{k}}$ 
(Fig.~\ref{fig:dd03}c).
This is because the models make different assumptions.
Here we assume that the size distribution is continuous between $D_{\rm{c}}$ and 
$D_{\rm{bl}}$,
whereas in DD03 the large planetesimals feeding the cascade are seen as separate from
the cascade.
This means that for us $M_{\rm{tot}}$ gives a direct estimate of $K$ (equation 
\ref{eq:mtot1})
and so the amount of dust $f$, while for DD03 they equate the mass flow through the cascade
with the mass input from the break-up of planetesimals meaning that while their scaling
parameter is proportional to $M_{\rm{tot}}$ (as is ours), it also includes a dependence on 
the
parameter we call $X_{\rm{c}}$ which affects the mass flow rate in the cascade.
This explains all of the differences:
the details of the scaling explain the slightly different initial $f$ values in all
the figures,
and the fact that for us planetesimals of size $D_{\rm{c}}$ are destroyed by planetesimals
down to size $X_{\rm{c}}D_{\rm{c}}$ means that our collisional lifetimes are always shorter 
than 
those in
DD03, since they assume that planetesimals only collide with same size planetesimals.
For us changing $v_{\rm{rel}}/v_{\rm{k}}$ does not affect the initial $f$ parameter as 
described
above, but it does affect the collisional lifetime of the largest planetesimals
which can survive longer if $v_{\rm{rel}}/v_{\rm{k}}$ is reduced (since this means that fewer
planetesimals in the cascade cause destruction on impact).
The opposite is the case for the DD03 model: changing $v_{\rm{rel}}/v_{\rm{k}}$ does not
affect the collisional lifetime of the largest planetesimals, since they only collide
with each other, but a lower collision velocity does increase the initial dust
luminosity because the cascade must have more mass in it to result in a mass flow
rate sufficient to remove mass introduced by the large planetesimals.
While the difference is subtle, it is important, since $v_{\rm{rel}}/v_{\rm{k}}$ may be 
important
in determining the presence of dust at late times (DD03; section \ref{s:hot}).

On the face of it, it seems that our model provides a more accurate description of the disk.
The reason is that in a collisional cascade the mass flow does not need to be taken into
account, since it results in the $q=11/6$ size distribution (Tanaka et al. 1996).
In other words the dependence of the scaling of the cascade with $X_{\rm{c}}$ found by
DD03 should have been removed if the largest planetesimals had been allowed to
collide with smaller planetesimals (since increasing $X_{\rm{c}}$ would have both
restricted mass flow within the cascade and slowed down the mass input from
the destruction of large planetesimals).
However, it is also true that the $q=11/6$ distribution only applies in an
infinite cascade, and since both models have truncated the size distribution
at $D_{\rm{c}}$, this would affect the evolution.
Also, the effect of the variation of $Q_{\rm{D}}^\star$ with $D$ on the size distribution
and its evolution are not yet clear, and neither is the evolution of the size 
distribution while the collisional cascade is being set up.
These issues will be discussed only briefly in this paper, in which the simple
evolution model described above is applied to some of the latest observational
results on debris disks.

\section{Application to rare systems with hot dust}
\label{s:hot}

\begin{deluxetable*}{cccccccc}
  \tabletypesize{\scriptsize}
  \tablecaption{Main sequence sun-like (F, G and K) stars in the literature
    with evidence for hot dust at $<10$ AU.
  \label{tab:hot} }
  \tablewidth{0pt}
  \tablehead{
  \colhead{Star name} & \colhead{Sp. Type} & \colhead{Age, Myr} & \colhead{Radius, AU} &
  \colhead{$f_{\rm{obs}} = L_{\rm{ir}} / L_\star $} & \colhead{$f_{\rm{max}}$} &
  \colhead{Transient?} & \colhead{Reference} }
  \startdata
    HD98800\tablenotemark{c}   & K4/5V & $\sim$ 10 & 2.2       & $220\times10^{-3}$ &  
    $270 \times 10^{-6}$  & Not req & Low et al. (2005) \\
    HD113766\tablenotemark{ac}  & F3V   & 16        & 3         & $2.1\times10^{-3}$ &  
    $45 \times 10^{-6}$  & Not req & Chen et al. (2005) \\
    HD12039         & G3/5V & 30        & 4-6       & $0.1\times10^{-3}$ &
    $200 \times 10^{-6}$  & Not req & Hines et al. (2006) \\ 
    BD+20307\tablenotemark{a}  & G0V   & 300       & 1         & $40\times10^{-3}$ &
    $0.36 \times 10^{-6}$  & Yes & Song et al. (2005) \\  
    HD72905\tablenotemark{a}   & G1.5V & 400       & 0.23\tablenotemark{b}  & $0.1\times10^{-3}$    & 
    $0.011 \times 10^{-6}$ & Yes & Beichman et al. (2006a) \\ 
    $\eta$ Corvi\tablenotemark{a}  & F2V   & 1000      & 1-2\tablenotemark{b}  & $0.5\times10^{-3}$ &  
    $0.15 \times 10^{-6}$  & Yes & Wyatt et al. (2005) \\ 
    HD69830\tablenotemark{a}   & K0V   & 2000      & 1         & $0.2\times10^{-3}$ &
    $0.13 \times 10^{-6}$ & Yes & Beichman et al. (2005)
  \enddata
  \tablenotetext{a}{infrared silicate feature}
  \tablenotetext{b}{also has cool dust component at $>10$ AU}
  \tablenotetext{c}{binary star}
\end{deluxetable*}

Very few main sequence stars exhibit hot dust within $\sim 10$ AU,
i.e., in the region where we expect planets may have formed.
Four surveys have searched for hot dust around sun-like stars
(main sequence F, G or K stars) by looking for a 25 $\mu$m flux in
excess of photospheric levels using IRAS (Gaidos 1999), ISO (Laureijs
et al. 2002) and Spitzer (Hines et al. 2006; Bryden et al. 2006).
All concluded that only $2 \pm 2$\% of these stars have hot dust with
infrared luminosities $f=L_{\rm{ir}}/L_\star > 10^{-4}$, finding a total
of 3 candidates.
Other hot dust candidates exist in the literature, however some IRAS
excess fluxes have turned out to arise from chance alignments with background 
objects (e.g., Lisse et al. 2002), including the candidate HD128400 from
the hot dust survey of Gaidos (1999) (Zuckerman, priv. comm.).
Thus confirmation of the presence of dust centred on the star using
ground- and space-based mid-IR imaging is vitally important (Smith, Wyatt
\& Dent, in prep.).
The tally of confirmed hot dust sources now stands at seven, and these
are summarized in Table \ref{tab:hot} which also gives the estimated
radial location of the dust based on fitting of the spectral energy
distribution of the excess emission;
for all stars the dust is predicted to lie at $<10$ AU.

While the frequency of the presence of such emission is low, there is
as yet no adequate explanation for its origin and why it occurs in so few 
systems.
Analogy with the solar system suggests that these are systems in which
we are witnessing the collisional grinding down of atypically massive
asteroid belts.
However, other scenarios have also been proposed in which the dust
is transient, having been produced in some stochastic process.
Such a process could be a recent collision between two massive
protoplanets in an asteroid belt (Song et al. 2005), the sublimation
of one supercomet (Beichman et al. 2005), or the sublimation of a swarm
of comets, possibly scattered in from several tens of AU in an
episode analogous to the period of Late Heavy Bombardment in the
solar system (Gomes et al. 2005).

\subsection{Are these massive asteroid belts?}
\label{ss:qe}

\begin{figure*}
  \centering
  \begin{tabular}{cc}
     \hspace{-0.15in} \includegraphics[width=3.2in]{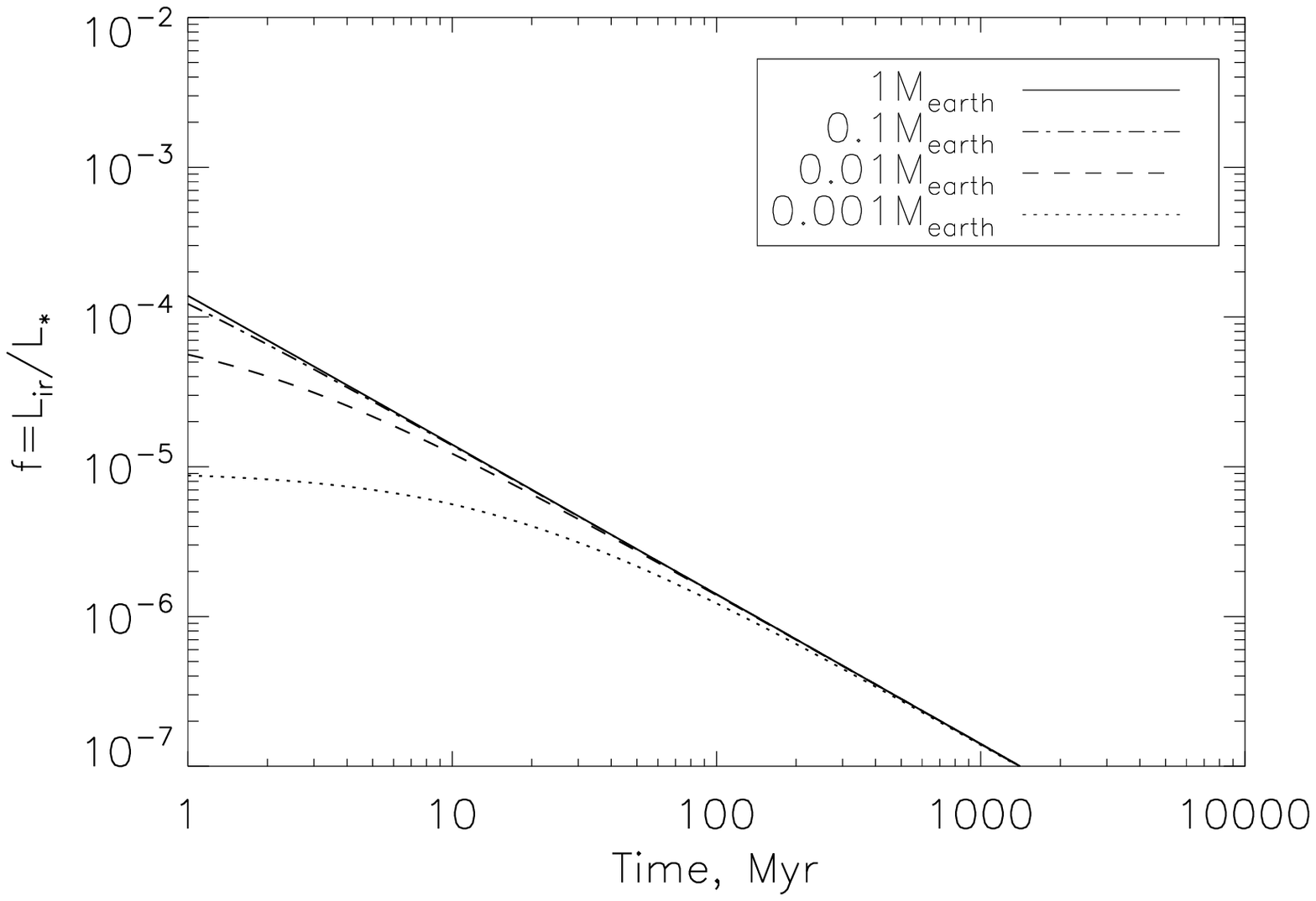} &
     \hspace{-0.15in} \includegraphics[width=3.2in]{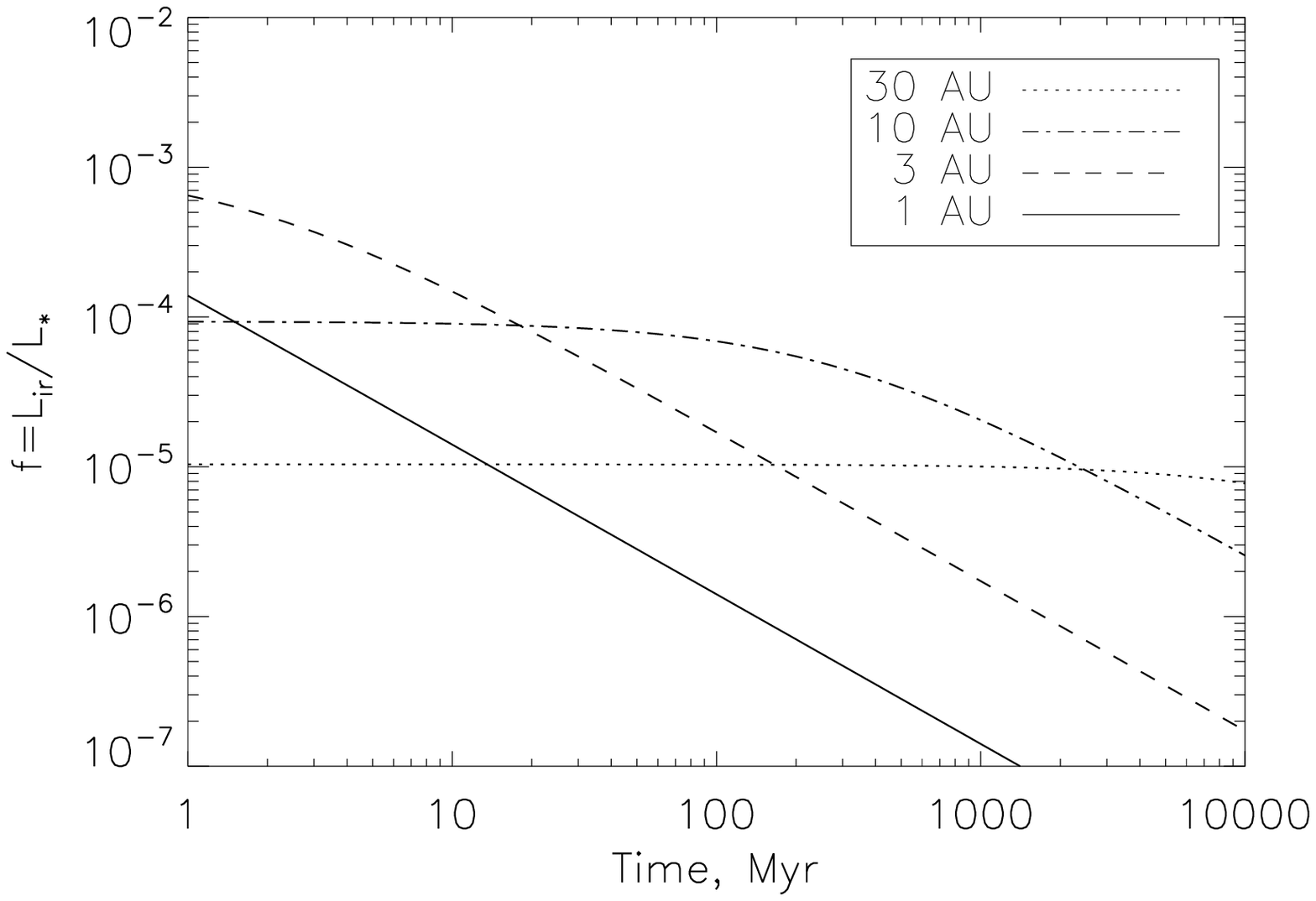} \\[-0.0in]
     \hspace{-0.15in} \includegraphics[width=3.2in]{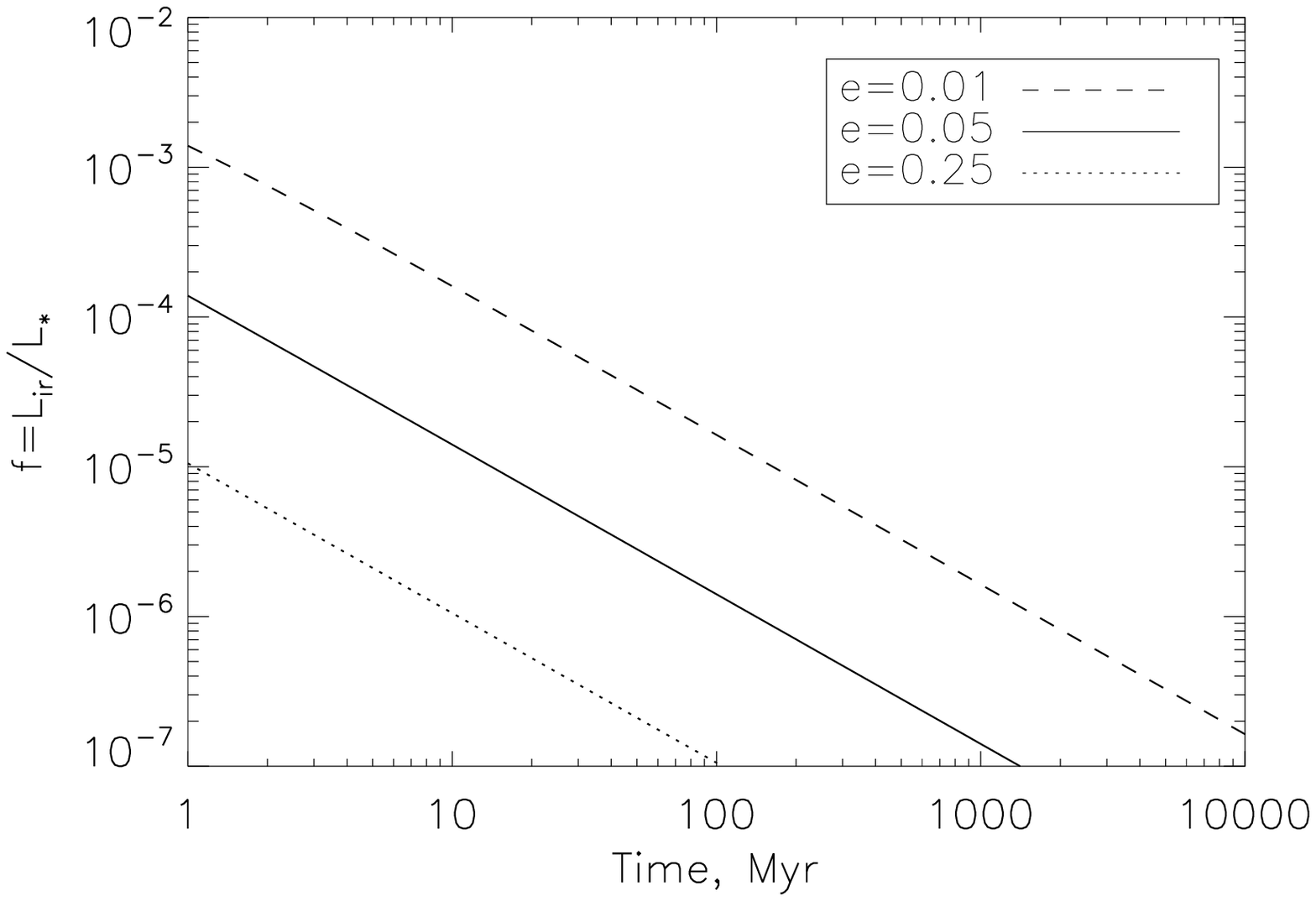} &
     \hspace{-0.15in} \includegraphics[width=3.2in]{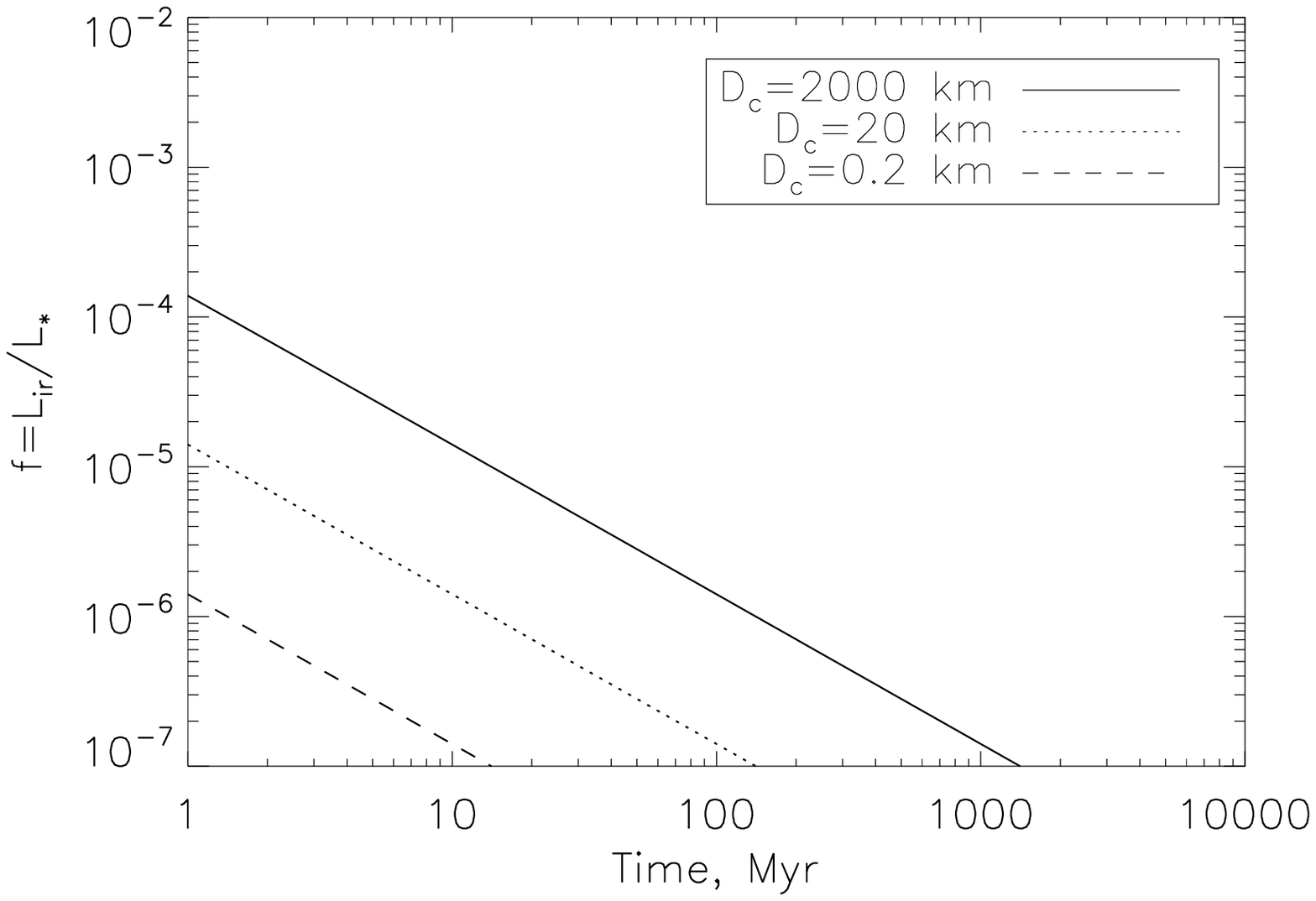} \\[-0.0in]
     \hspace{-0.15in} \includegraphics[width=3.2in]{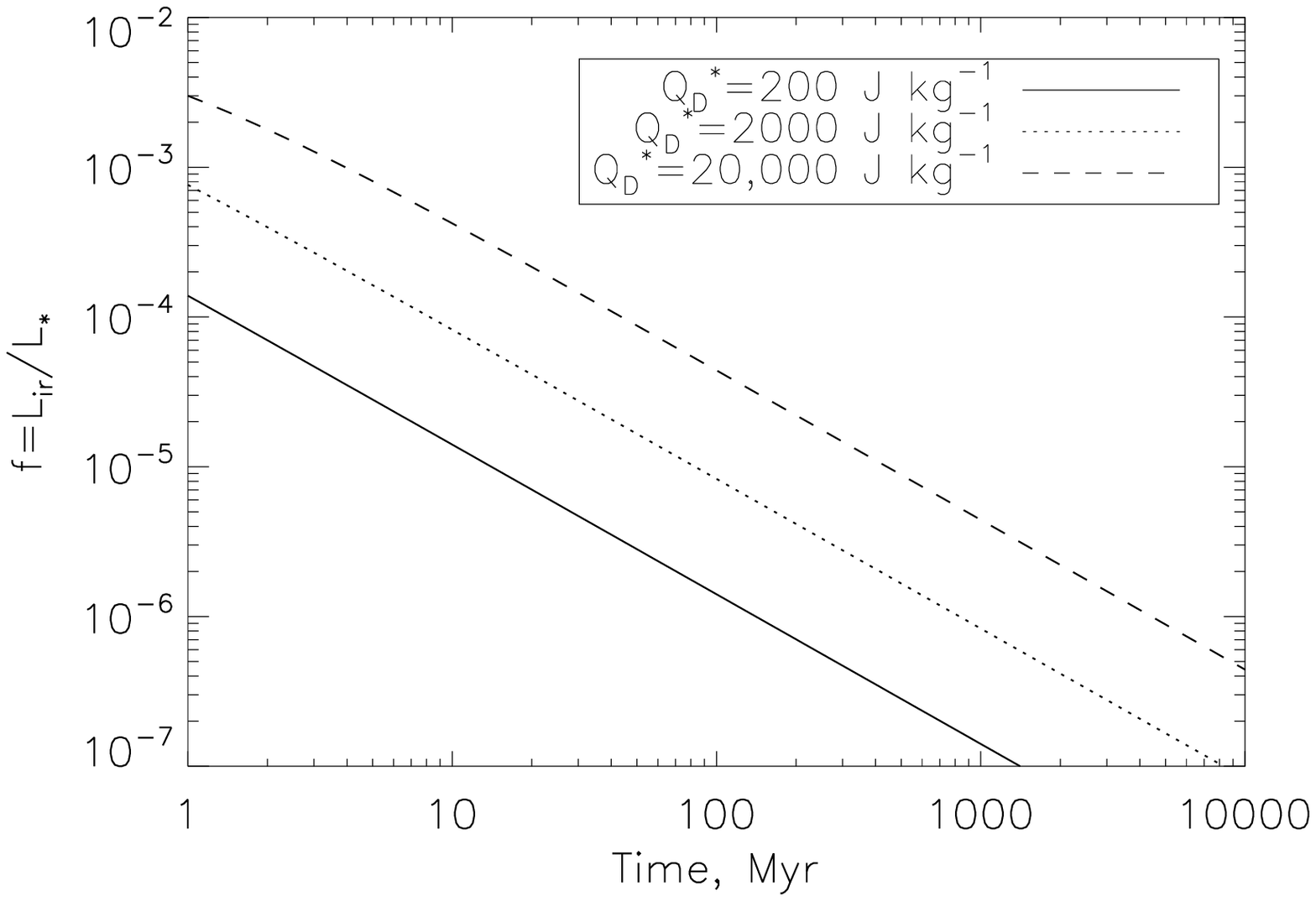} &
     \hspace{-0.15in} \includegraphics[width=3.2in]{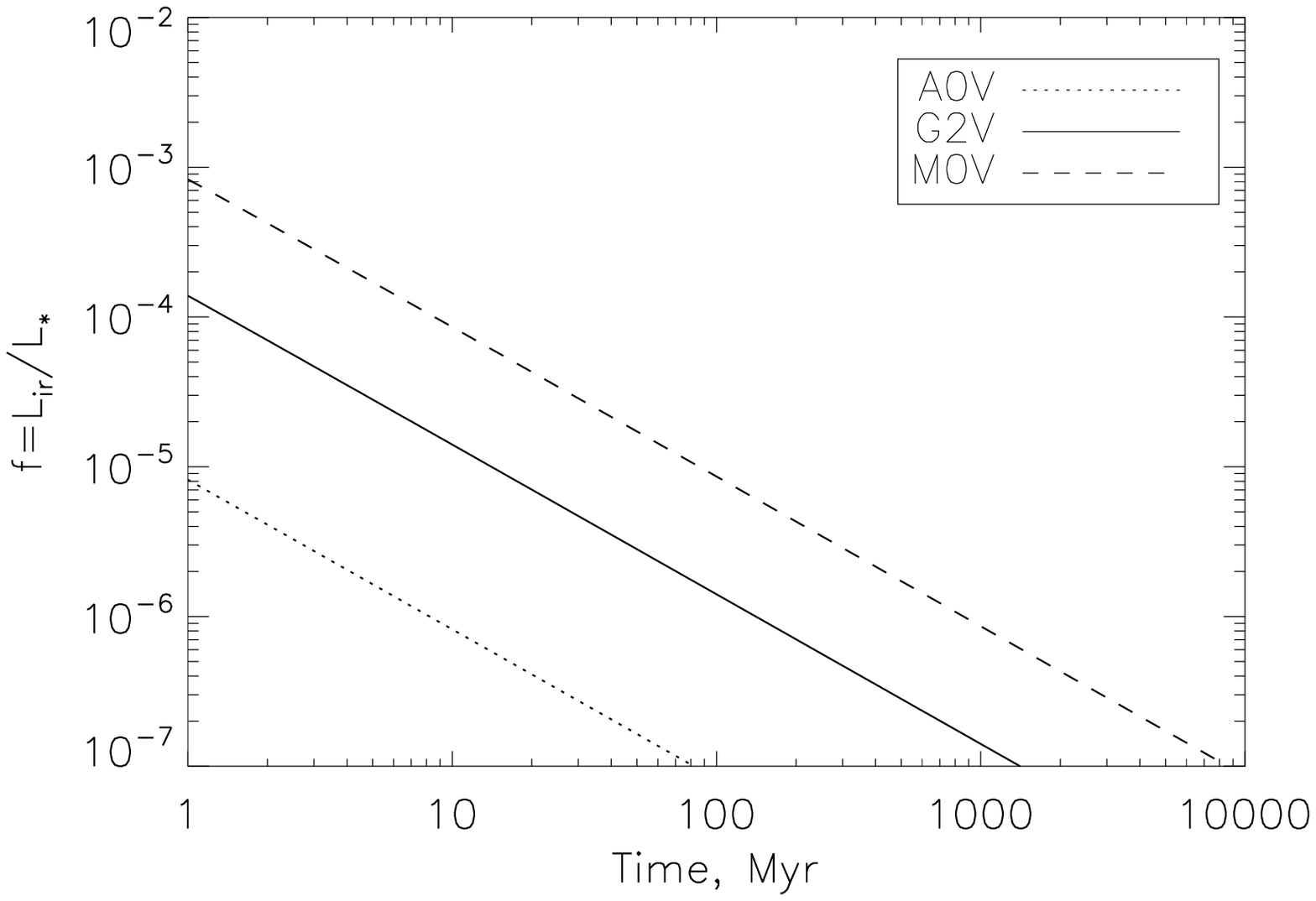}
  \end{tabular}
  \caption{The collisional evolution of a planetesimal belt at $r=1$ AU around a sun-like
  star ($L_\star=M_\star=1$) of initial mass $M_{\rm{tot}}(0)=1M_\oplus$ assuming that belt
  can be described by the parameters used in \S \ref{ss:qe} (i.e., $dr/r=0.5$,
  $D_{\rm{c}}=2000$ km, $\rho=2700$ kg m$^{-3}$, $e=0.05$, $e/I=1$,
  $Q_{\rm{D}}^\star=200$ J kg$^{-1}$).
  All panels show dust luminosity $f=L_{\rm{ir}}/L_\star$ as a function of time, and
  the evolution with the above nominal parameters is shown with a solid line.
  The different panels show the effect of changing the following parameters:
  \textbf{(top left)} starting disk mass $M_{\rm{tot}}(0)$,
  \textbf{(top right)} disk radius $r$,
  \textbf{(middle left)} planetesimal eccentricity $e$,
  \textbf{(middle right)} maximum planetesimal size $D_{\rm{c}}$,
  \textbf{(bottom left)} planetesimal strength $Q_{\rm{D}}^\star$, and
  \textbf{(bottom right)} stellar spectral type.}
  \label{fig:hotevol}
\end{figure*}

Here we consider the possibility that these are atypically massive
asteroid belts, and show that for the majority of the known systems this
is unlikely to be the case.
The reason is that given in \S \ref{ss:ce}, which is that more massive
asteroid belts are not necessarily more dusty at late times, and there
is a maximum dust luminosity we can expect for a belt of a given age,
given its radial location (equations \ref{eq:mmax1}-\ref{eq:fmax3}).
To arrive at a rough estimate of the maximum possible $f_{\rm{max}}$
we assume the following parameters:
the largest possible planetesimal is $D_{\rm{c}} = 2000$ km, since this is above the
largest members of the asteroid and Kuiper belts, and fits with the
expectation that planetesimal growth is halted once the largest planetesimals
reach this size due to the resulting gravitational perturbations (Kenyon \& Bromley 
2002);
belt width is $dr = 0.5r$;
planetesimal strength is $Q_{\rm{D}}^\star = 200$ J kg$^{-1}$, the canonical
value used in DD03, although gravity strengthening can give rise to higher
values for planetesimals larger than $\sim 1$ km (see \S \ref{ss:pc});
planetesimal eccentricity is $e = 0.05$, typical for planetesimal belts
like the asteroid belt that are undergoing a collisional cascade, and close
to that expected from stirring by 2000 km planetesimals within such belt.
\footnote{Equating the velocity dispersion in the belt with the escape
velocity of a planetesimal of size $D_{\rm{c}}$ gives
$e \approx 2.6 \times 10^{-7} \rho^{0.5} r^{0.5} M_\star^{-0.5} D_{\rm{c}}$. }
Substituting in these nominal values into equation (\ref{eq:fmax3}) and
approximating $M_\star=L_\star=1$ gives:
\begin{equation}
  f_{\rm{max}} = 0.16 \times 10^{-3} r^{7/3} t_{\rm{age}}^{-1}. \label{eq:fmax4}
\end{equation}
Plots analogous to those in Fig.~\ref{fig:dd03} are presented in Fig.~\ref{fig:hotevol}
which shows the evolution for a planetesimal belt with the nominal parameters
described above (and with a nominal starting mass of $M_{\rm{tot}}(0)=1M_\oplus$) along
with the consequence for the evolution of changing any of those parameters.
Note that it is most appropriate to refer to Fig.~\ref{fig:hotevol}, rather than
Fig.~\ref{fig:dd03}, when considering the evolution of planetesimal belts close to
sun-like stars.

The value of $f_{\rm{max}}$ is quoted in Table \ref{tab:hot} under the assumption
that the planetesimal belt has the same age as the star.
The quoted value for each star is that from equation (\ref{eq:fmax2}) for its
spectral type, but is within a factor of three of that given in equation
(\ref{eq:fmax4}), indicating that this equation may be readily applied to
observed belts in the future.
The four oldest systems (BD+20307, HD72905, $\eta$ Corvi and HD69830)
have $f_{\rm{obs}} \gg 10^{3}f_{\rm{max}}$.
We show in \S \ref{ss:pc} that even with a change in parameters it is not
possible to devise asteroid belts in these systems that could survive to the age of
the stars giving rise to the observed dust luminosities.
Thus we conclude that this period of high dust luminosity started
relatively recently.
The timescale over which a belt can last above a given
luminosity, $f_{\rm{obs}}$, is $t_{\rm{age}}f_{\rm{max}}/f_{\rm{obs}}$,
since collisions would grind a belt down to this level on such a timescale.
This implies that belts this luminous only last between a few thousand years
(BD+20307 and HD72905) and a few Myr ($\eta$ Corvi and HD69830).
However, the true duration of this level of dust luminosity depends
on the details of the process causing it, and moreover there is still
up to two orders of magnitude uncertainty in $f_{\rm{max}}$ (see \S \ref{ss:pc}).
Thus this calculation should not yet be used to infer from the $\sim 2$\% of
systems with hot dust that, e.g., every sun-like star must undergo 10-1000 such
events in its lifetime (or fewer systems must undergo even more events).
For now the conclusion is that these systems cannot be planetesimal
belts that have been evolving in a collisional cascade for the full age of the
star.

This leaves open the possibility that the collisional cascade in these systems
was initiated much more recently, perhaps because a long timescale was required
to form the 2000-3000 km sized planetesimals necessary to stir the planetesimal
belt and cause the switch from accretion to collisional cascade (Kenyon \& Bromley
2004).
However, we consider this to be unlikely, because the timescale for the formation
of objects of this size at 1 AU from a solar mass star was given in
Kenyon \& Bromley (2004) to be $\sim 0.6 dr/M_{\rm{tot}}$ Myr, where
$M_{\rm{tot}}$ is the mass of material in an annulus of width
$dr$, just as in the rest of the paper.
This means that the cascade can only be delayed for 100-1000 Myr at 1 AU for planetesimal
belts of very low mass, which would also be expected to have low dust luminosities when
the cascade was eventually ignited.
For example, a delay of $>500$ Myr would require $<0.6 \times 10^{-3}M_\oplus$ in the
annulus at 1 AU of 0.5 AU width, a mass which corresponds to a fractional luminosity
of $<5 \times 10^{-6}$ (equations \ref{eq:mtot2} and \ref{eq:f} with $\rho = 2700$ kg 
m$^{-3}$ and $q=11/6$), much lower than that observed in all systems.
One can also consider the same argument in the following way:
the observed luminosity $f_{\rm{obs}}$ implies a planetesimal belt mass which current
planet formation theories indicate would result in the growth of 2000 km planetesimals
which would ignite a collisional cascade on a timescale of
$3 \times 10^{-3}(dr/r)/f_{\rm{obs}}$ Myr if this was placed at 1 AU from a solar mass star.
The conclusion at the end of the last paragraph also considers the collisional cascade to evolve 
in quasi-steady state, and it is possible that collisions between large members of the cascade 
may have recently introduced large quantitites of small dust;
that possibility is discussed in \S \ref{ss:singlecoll}.

For the three youngest systems the conclusions are less clear.
The dust luminosities of HD12039 and HD113766 are, respectively, close to
and fifty times higher than the maximum allowed value for collisionally evolved 
planetesimal belts.
However, given the uncertainties in the parameters in the model (described in \S 
\ref{ss:pc}), we conclude that it is not possible to say that these could not be massive 
asteroid belts.
The main reason that firm conclusions cannot be drawn is the large radial location
of the dust at $>2$ AU.
The strong dependence of $f_{\rm{max}}$ on $r$ means that it is easiest to
constrain the nature of belts within a few AU which evolve very rapidly.
For the youngest system (HD98800), while its dust luminosity lies a factor of 
800 above the maximum for the age of the star, we do not infer
that this must be transient, since the high dust luminosity and low age imply
that this system is in a transitional phase and the collisional cascade in
this debris disk is likely to have only recently been ignited.
Rather we note that this model implies that due to collisional processing
this debris disk cannot maintain this level of dust emission beyond the
next $\sim 10,000$ years (albeit with an additional two orders of magnitude
uncertainty, \S \ref{ss:pc}).

\subsection{Possible caveats}
\label{ss:pc}
Given the large number of assumptions that went into the estimate for
$f_{\rm{max}}$, it is worth pointing out that this model is in excellent
agreement with the properties of the asteroid belt in the solar system,
since for a 4500 Myr belt at 3 AU the model predicts
$M_{\rm{max}}=0.4 \times 10^{-3}M_\oplus$, which is close to the inferred
mass of the asteroid belt of $0.6 \times 10^{-3}M_\oplus$ (Krasinsky et al. 2002).
The model also predicts $f_{\rm{max}}=5 \times 10^{-7}$, which is consistent
with the estimate for the zodiacal cloud of
$L_{\rm{ir}}/L_\star = 0.8 \times 10^{-7}$ (Backman \& Paresce 1993).
\footnote{In planetesimal belts as tenuous as the asteroid belt, the
effect of P-R drag is important (Wyatt 2005) meaning that the cross-sectional
area of dust in the zodiacal cloud is dominated by $\sim 100$ $\mu$m sized grains
rather than grains of size $D_{\rm{bl}}$ as assumed in the simple model of
\S \ref{ss:pb}.
Taking this into account would reduce the fractional luminosity predicted by the model
by an order of magnitude.}
It is also necessary to explore if there is any way in which the parameters
of the model could be relaxed to increase $f_{\rm{max}}$ and so change the
conclusions about the transience of the hot dust systems.
Equation (\ref{eq:fmax3}) indicates one way in which
$f_{\rm{max}}$ could be increased, which is by either reducing the eccentricities of the 
planetesimals, $e$, or increasing their strength, $Q_{\rm{D}}^\star$, both of which could 
increase $X_{\rm{c}}$ and so decrease the rate at which mass is lost from the cascade
(e.g., fig.~\ref{fig:hotevol}).
The other way is to change the size distribution so that a given disk mass results
in a significantly larger dust luminosity, e.g., by increasing $q$.

In fact Benz \& Asphaug (1999) found a value of $Q_{\rm{D}}^\star$ that is higher 
than $2 \times 10^{5}$ J kg$^{-1}$ for planetesimals as large as 2000 km for both ice
and basalt compositions.
This would result in an increase in $f_{\rm{max}}$ by a factor of $\sim 170$
(e.g., Fig.~\ref{fig:hotevol}).
However, such a high value of $Q_{\rm{D}}^\star$ is possible only due to gravity 
strengthening
of large planetesimals, and the dependence in this regime of $Q_{\rm{D}}^\star \propto 
D^{1.3}$ 
(Benz \& Asphaug 1999) would result in an equilibrium size distribution with
$q_{\rm{g}} \approx 1.68$, since when $Q_{\rm{D}}^\star \propto D^s$ then $q = (11+s)/(6+s)$
(O'Brien \& Greenberg 2003).
If such a distribution was to hold down to the smallest dust grains the net result
would be a decrease in $f_{\rm{max}}$ by $\sim 200$.
This is not the case, however, since objects in the size
range $D<D_{\rm{t}}=0.15$ km are in the strength scaled regime where $Q_{\rm{D}}^\star \propto 
D^{-0.4}$
leading to a size distribution with $q_{\rm{s}} = 1.89$ in this range.
According to O'Brien \& Greenberg (2003) the size distribution of a
collisional cascade with a realistic $Q_{\rm{D}}^\star$ prescription
should have two components (characterized by $q_{\rm{g}}$ and $q_{\rm{s}}$), but there is a 
discontinuity at the transition size $D_{\rm{t}}$ with the strength scaled component
shifted down by an appropriate amount $x_{\rm{t}}$ (see their Fig. 3b).
This means that $f_{\rm{max}}$ should be higher than that derived using
equation (\ref{eq:fmax1}) with $q=q_{\rm{g}}$ by a factor
$x_{\rm{t}} 
(3q_{\rm{g}}-5)(3q_{\rm{s}}-5)^{-1}(D_{\rm{bl}}/D_{\rm{t}})^{3(q_{\rm{g}}-q_{\rm{s}})}$.
Since $x_{\rm{t}}<1$, then substituting the values from Benz \& Asphaug (1999)
given above implies that Table \ref{tab:hot} underestimates $f_{\rm{max}}$ by at
most a factor of 50-100 (possibly much less).
In other words, we anticipate that by including a more realistic prescription for 
$Q_{\rm{D}}^\star$ and the resulting size distribution, this would change the inferred 
$f_{\rm{max}}$ but not upwards by an amount more than two orders of magnitude.
For this reason, transience is only inferred for those systems for which
$f_{\rm{obs}}/f_{\rm{max}} \gg 100$.

A lower eccentricity is, however, one potential avenue for increasing the amount
of dust remaining at late times.
Equation (\ref{eq:fmax3}) shows that since $G(11/6,X_{\rm{c}}) \propto e^{5/3}$
(Fig.~\ref{fig:gvse}a), this means that reducing $e$ from 0.05 to 0.01 or 0.001 gives
a decrease in $G(11/6,X_{\rm{c}})$ of 15 or 680, and so an increase in $f_{\rm{max}}$
by a corresponding amount (e.g., Fig.~\ref{fig:hotevol}).
In fact the increase can be much more than this, since when $e$ is reduced to levels
below $4.7 \times 10^{-5} \sqrt{Q_{\rm{D}}^\star rM_\star^{-1}[1.25+(I/e)^2]^{-1}}$ 
then $X_{\rm{c}}>1$ (e.g., Fig.~\ref{fig:gvse}b).
In such a regime mutual collisions do not result in the destruction of 
planetesimals, rather in their merger and growth.
At this point $G(11/6,X_{\rm{c}})<0$, i.e., $f_{\rm{max}}$ is infinite since, in this
simple model, whatever the starting conditions there is no evolution (although
in practice the size distribution would evolve due to planetesimal growth).
At $\sim 1$ AU, this means $e$ must be larger than
0.0005 (for $Q_{\rm{D}}^\star = 200$ J kg$^{-1}$, appropriate for $D_{\rm{c}}=0.15$ km)
or 0.014 (for $Q_{\rm{D}}^\star = 2 \times 10^5$ J kg$^{-1}$, appropriate for $D_{\rm{c}}=2000$ km)
to initiate a collisional cascade, values which are consistent with those quoted by
more detailed planet formation models (e.g., Kenyon \& Bromley 2002).
Such eccentricities would be expected through stirring either by $>1000$ km
planetesimals which formed within the belt, or by more massive perturbers which formed 
outside the belt, both of which can be expected to occur within 10-100 Myr
(Kenyon \& Bromley 2006).
This was considered in \S \ref{ss:qe} where it was shown that the cascade would
be initiated following the growth of $\sim 2000$ km planetesimals on timescales
that are much shorter than the age of the system for the disk masses required to
produce a dust luminosity at the observed level.

The only route which could plausibly maintain the hot dust systems in Table \ref{tab:hot}
in collisional equilibrium over the age of the stars might be to invoke some mechanism
which maintains the eccentricity at a level which the cascade is only just being eroded.
However, Fig.~\ref{fig:gvse}a shows that $G(11/6,X_{\rm{c}})$ is a strong function
of $e$ when $G(11/6,X_{\rm{c}}) < 1$, since the range $G(11/6,X_{\rm{c}})=0-1$ is
covered by a factor of less than two in eccentricity.
Thus we consider it reasonable to assume that the best possible combination of 
$Q_{\rm{D}}^\star$ and $e$ in this regard would result in 
$G(11/6,X_{\rm{c}}) \approx 1$ (corresponding to $X_{\rm{c}}=0.69$);
lower values of $G(11/6,X_{\rm{c}})$ are possible, but only within a very narrow
range of eccentricity.
Since in the above example with a realistic $Q_{\rm{D}}^\star$ prescription extending
up to 2000 km we assumed $e=0.05$ which already resulted in $G(q_{\rm{g}},X_{\rm{c}}) < 1$, 
we consider that it is not reasonable to fine tune the eccentricity further 
to increase $f_{\rm{max}}$;
e.g., decreasing to $e=0.03$ results in some disks not evolving and the rest
with $f_{\rm{max}}$ higher than that quoted in Table \ref{tab:hot} by a factor $\sim
150$.
Thus we conclude that the estimate given in Table \ref{tab:hot} (and e.g., equation
\ref{eq:fmax1}) underestimates $f_{\rm{max}}$ by at most a factor of $\sim 100$, unless
the eccentricity happens to lie within $\pm 10\%$ of a critical value.

It is also worth noting that low levels of eccentricity would result in large 
gravitational focussing factors for large planetesimals which would enhance
$f_{\rm{cc}}$ and so decrease the time for these planetesimals to be catastrophically
destroyed, something which is compounded by the higher collision velocity
in gravitationally focussed collisions which reduces $X_{\rm{c}}$ because collisions
with smaller planetesimals can cause catastrophic disruption (e.g., equation
\ref{eq:xc2}).
However, we do not need to account for this here, since gravitational focussing
becomes important when $v_{\rm{rel}} < v_{\rm{esc}} \approx \sqrt{(2/3)\pi \rho G}(10^{-3}D)$
and so when $e<4\times 10^{-7}\sqrt{\rho r M_\star^{-1}[1.25+(I/e)^2]^{-1}}D$
(where $D$ is in km);
i.e., when $e<2 \times 10^{-6}$ for $D=0.15$ km and $e<0.027$ for $D=2000$ km
at 1 AU from a $1M_\odot$ star, both of which occur close to or below the level
at which collisions result in accumulation rather than destruction.

\subsection{Are these the products of single collisions?}
\label{ss:singlecoll}
One possible origin for the hot dust which is quoted in the literature is that
it is the product of a single collision (Song et al. 2005).
Our model can be used to make further predictions for the likelihood of massive collisions
occurring within an asteroid belt.
The maximum number of parent bodies (i.e., planetesimals) larger than $D_{\rm{pb}}$ 
remaining at late times occurs when $M_{\rm{tot}}=M_{\rm{max}}$ and so is given by:
\begin{eqnarray}
  n(D>D_{\rm{pb}}) & = & \left( \frac{5.6 \times 10^{10}r^{3.5} (dr/r)}
                                     {M_\star^{0.5} D_{\rm{c}}^2 t_{\rm{age}}} \right)
            \left[ \left( \frac{D_{\rm{c}}}{D_{\rm{pb}}} \right)^{3q-3}-1  \right]
            \nonumber \\
  & & \times \left( \frac{(3q-5)[1+1.25(e/I)^2]^{-0.5}}{(3-3q)G(q,X_{\rm{c}})} \right).
            \label{eq:ndgtdpb}
\end{eqnarray}
The collision timescale for planetesimals of size $D_{\rm{pb}}$ is
\begin{eqnarray}
  t_{\rm{c}}(D_{\rm{pb}}) & = & t_{\rm{c}}(D_{\rm{c}}) 
                                f_{\rm{cc}}(D_{\rm{c}})/f_{\rm{cc}}(D_{\rm{pb}}) \nonumber \\ 
                          & = & 10^6 t_{\rm{age}}(D_{\rm{pb}}/D_{\rm{c}})^{3q-5}, 
\label{eq:tcd}
\end{eqnarray}
noting that the collisional lifetime of the largest planetesimals, $t_{\rm{c}}(D_{\rm{c}})$, 
is the age of the star for a planetesimal belt at maximum luminosity for this
age.
These can be combined to give the destructive collision rate for planetesimals larger
than $D_{\rm{pb}}$:
\begin{eqnarray}
  dN_{\rm{c}}(D>D_{\rm{pb}})/dt & = & 1000r^{13/3}(dr/r)t_{\rm{age}}^{-2}D_{\rm{c}}
                                      D_{\rm{pb}}^{-3}\times \nonumber \\ 
                    & &  M_\star^{-4/3}{Q_{\rm{D}}^\star}^{5/6}e^{-5/3},
  \label{eq:dncdt}
\end{eqnarray}
in Myr$^{-1}$, where the assumptions that $q=11/6$, $e=I$ and $X_{\rm{c}} \ll 1$ have been
used in deriving this equation.

We now assume that we are considering collisions capable of reproducing the
observed dust level, $f_{\rm{obs}}$, so that the lifetime of the resulting collision
products can be estimated from the collisional lifetime of that dust, assumed to be
of size $D_{\rm{bl}}$ (WD02):
\begin{equation}
  t_{\rm{c}}(D_{\rm{bl}}) = 0.04 r^{1.5}M_\star^{-0.5}(dr/r)f_{\rm{obs}}^{-1}, \label{eq:tcdust}
\end{equation}
in years, noting that collisions would remove the dust on a faster timescale than P-R drag
(Wyatt 2005; Beichman et al. 2005).
Combining equations (\ref{eq:dncdt}) and (\ref{eq:tcdust}) gives the fraction
of time that collisions are expected to result in dust above a given level of
$f_{\rm{obs}}$:
\begin{eqnarray}
  P(f>f_{\rm{obs}}) & = & 4 \times 10^{-5}r^{35/6}(dr/r)^2t_{\rm{age}}^{-2}
                 D_{\rm{c}}D_{\rm{pb}}^{-3} \times \nonumber \\
               & & M_\star^{-11/6}f_{\rm{obs}}^{-1}
                 {Q_{\rm{D}}^\star}^{5/6}e^{-5/3}.
  \label{eq:pffobs}
\end{eqnarray}

To estimate the minimum size of the parent body, $D_{\rm{pb}}$, responsible for this
dust, we consider how large a planetesimal must be to reproduce $f_{\rm{obs}}$ if a
destructive collision resulted in one fragment with half the mass of the original
planetesimal (i.e., the definition of a destructive collision), with the remaining
mass in particles of size $D_{\rm{bl}}$:
\begin{equation}
  D_{\rm{pb}} = 890[D_{\rm{bl}}r^2f_{\rm{obs}}]^{1/3}. \label{eq:dpb}
\end{equation}

\begin{deluxetable*}{ccccccc}
  \tabletypesize{\scriptsize}
  \tablecaption{Parameters in the model for the hot dust systems of Table \ref{tab:hot}
  used to determine whether the observed dust can be the outcome of a single
  collision in a massive asteroid belt which is itself not normally bright
  enough to be detected.
  \label{tab:hot2} }
  \tablewidth{0pt}
  \tablehead{ \colhead{Star name} & \colhead{$D_{\rm{pb}}$, km}  &  \colhead{$N(D>D_{\rm{pb}})$}  & 
    \colhead{$dN_{\rm{c}}(D>D_{\rm{pb}})/dt$, Myr$^{-1}$}  &
    \colhead{$t_{\rm{c}}(D_{\rm{bl}})$, yr}  &  
     \colhead{$P(f>f_{\rm{obs}})$}  &  \colhead{Single collision?} }
  \startdata
    HD98800   &  530  &  200          & 41      & 0.36 & $15 \times 10^{-6}$   & No \\
    HD113766  &  280  &  890          & 150     & 41   & $6100 \times 10^{-6}$ & Not imposs \\
    HD12039   &  110  &  77,000      & 12,000  & 2300 & 27\tablenotemark{*}     & Not imposs \\
    BD+20307  &  320  &  0.47\tablenotemark{*} & 0.0039  & 0.49 & $0.0019 \times 10^{-6}$  & No \\
    HD72905   &  15   &  1.4          & 0.036   & 22   & $0.79 \times 10^{-6}$  & No \\
    $\eta$ Corvi & 110&  7.8          & 0.033   & 59   & $2.0 \times 10^{-6}$  & No \\
    HD69830   &  39   &  19           & 0.068   & 110  & $7.7 \times 10^{-6}$ & No
  \enddata
  \tablenotetext{*}{For disks with $P(f>f_{\rm{obs}})>1$, this value indicates the number of
  collisions at that level we can expect to see in the disk at any one time.
  Likewise, for disks with $N(D>D_{\rm{pb}})<1$, this value indicates the probability
  that there is an object of this size remaining in the disk.}
\end{deluxetable*}

Table \ref{tab:hot2} lists the parameters for the hot dust systems assuming the
canonical parameters of $Q_{\rm{D}}^\star=200$ J kg$^{-1}$, $D_{\rm{c}}=2000$ km and 
$e=0.05$.
To determine whether a system could have been reproduced by a single collision, the
final value of $P(f>f_{\rm{obs}})$ was compared with the statistic that 2\% of systems
exhibit hot dust (which therefore considers the optimistic case where all stars
have planetesimal belts at a few AU).
For the systems which were inferred in Table \ref{tab:hot} to be transient,
all are extremely unlikely ($<0.001$\%) to have been caused by a single collision amongst 
planetesimals in a planetesimal belt which has undergone a collisional cascade since the star
was born.

While this statistic is subject to the uncertainties in the model parameters described
in \S \ref{ss:pc}, and so could be in error by around two orders of magnitude,
it must also be remembered that the most optimistic assumptions were used to arrive at
this figure.
For example, it is unlikely that the destruction of planetesimals of size $D_{\rm{pb}}$ would
release half of the mass of the planetesimal into dust $D_{\rm{bl}}$ in size.
\footnote{Such an optimistic assumption should not be dismissed out of hand, however, since 
the large amount of collisional processing that must have taken place means that
planetesimals more than a few km would be rubble piles.
These would have undergone 
shattering and reaccumulation numerous times meaning that they could have deep dusty
regolith layers which could be preferentially ejected in a collision.}
On the other hand, one might consider that the lifetime of the observed dust, 
$t_{\rm{c}}(D_{\rm{bl}})$, is an
underestimate of the duration of dust at the level of $f>f_{\rm{obs}}$, since the dust
could be replenished from the destruction of larger particles.
Indeed Farley et al. (2006) model the destruction of a 150 km planetesimal in the
asteroid belt and inferred a dust peak that lasted $\sim 1$ Myr, precisely because
large fragments produced in the collision replenished the dust population.
However, it should be cautioned that the dust peak inferred by Farley et al. (2006)
would not have been detectable as an infrared excess since it only caused a factor
$\sim 10$ enhancement in the luminosity of the zodiacal cloud (i.e., to
$f \approx 0.8 \times 10^{-6}$), and that in the context of our model, invoking a
population of larger grains that result from the collision would lead to a 
larger parent body (i.e., a larger $D_{\rm{pb}}$) required to reproduce the observed luminosity 
$f_{\rm{obs}}$ and so less frequent collisions;
i.e., it may be possible (even desirable) to increase $t_{\rm{c}}(D_{\rm{bl}})$, but only at the
expense of decreasing $dN_{\rm{c}}(D>D_{\rm{pb}})/dt$ leading to little change in 
$P(f>f_{\rm{obs}})$.
We note that $t_{\rm{c}}(D_{\rm{bl}})$ given in Table (\ref{tab:hot2}) is sufficiently short
that a measurement of the variability of the infrared excess on realistic (few year)
timescales could lead to constraints on the size of the grains feeding the
observed phenomenon, since if a population of larger grains existed then the
luminosity would fade on much longer timescales.

A further argument against the transient disks being caused by single collisions
is the fact that the probability of seeing the outcome of a collision, $P(f>f_{\rm{obs}})$,
falls off $\propto t_{\rm{age}}^{-2}$, which means that we would expect to see more
transient disks around younger stars than around older stars
(because young stars have more massive disks with more large
planetesimals and so more frequent collisions).
There is some evidence from Table \ref{tab:hot} that transience is 
more common around young systems, since none of the transient systems is older than
2 Gyr, whereas sun-like stars in the solar neighborhood would be expected to
have a mean age of $\sim 5$ Gyr.
However, while the statistics are poor, a $t_{\rm{age}}^{-2}$ dependence does seem
to be ruled out;
e.g., we would have expected to have detected 10 times more transient disks
caused by single collisions in the age range 50-500 Myr\footnote{It is not reasonable
to extend the age range to younger systems, since, as noted in \S \ref{ss:qe} it is
hard to discern whether or not dust detected in such systems is transient.}
than in the age range 0.5-5 Gyr, whereas 2 transient disks are known in the younger age bin, 
and 2 in the older age bin which is more consistent with a $t_{\rm{age}}^{-1}$ dependence.

\begin{deluxetable*}{ccccc}
  \tabletypesize{\scriptsize}
  \tablecaption{Parameters in the model for the transient hot dust systems of Table \ref{tab:hot}
  used to determine whether the observed dust could originate in the destruction of a
  planetesimal belt coincident with the dust:
  $dM_{\rm{loss}}/dt$ is the observed mass loss rate,
  $M_{\rm{max}}$ is the maximum mass of a planetesimal belt that is coincident with the dust given
  the age of the star,
  $t(f>f_{\rm{obs}})$ is the length of time such a planetesimal belt could sustain the observed
  dust luminosity,
  and $r_{\rm{out}}(\rm{100Myr})$ is the radius of a planetesimal belt which 
  would still have enough mass
  to sustain the observed dust luminosity for 100 Myr. \label{tab:hot3} }
  \tablewidth{0pt}
  \tablehead{ \colhead{Star name}    &
  \colhead{$dM_{\rm{loss}}/dt$, $10^{-6}M_\oplus$Myr$^{-1}$}  &
  \colhead{$M_{\rm{max}}$, $10^{-6}M_\oplus$} &
  \colhead{$t(f>f_{\rm{obs}})$, Myr} & 
  \colhead{$r_{\rm{out}}(\rm{100Myr})$, AU} }
  \startdata
    BD+20307     & $8.0 \times 10^6$ & 53    & $6.7 \times 10^{-6}$ & 45  \\
    HD72905      & 19                & 0.072 & $3.7 \times 10^{-3}$ & 2.4 \\
    $\eta$ Corvi & 2500              & 57    & 0.023                & 9.6 \\
    HD69830      & 64                & 12    & 0.18                 & 4.5
 \enddata
\end{deluxetable*}

In fact, within the context of this model, all of the disks which we infer to be
transient would also be inferred to not be the product of single
collisions.
This is evident by substituting $D_{\rm{pb}}$ from equation (\ref{eq:dpb}) and 
$f_{\rm{max}}$ from equation (\ref{eq:fmax2}) into equation (\ref{eq:pffobs}) to get:
\begin{equation}
  P(f>f_{\rm{obs}}) = 0.2 \times 10^6 (f_{\rm{max}}/f_{\rm{obs}})^2
                      (M_\star e^2 r^{-1} {Q_{\rm{D}}^\star}^{-1})^{5/6},
  \label{eq:pff2}
\end{equation}
which reduces to $P(f>f_{\rm{obs}})=16(f_{\rm{max}}/f_{\rm{obs}})^2(M_\star r^{-1})^{5/6}$
for the canonical parameters used before.
Since transient disks are defined by $f_{\rm{obs}}/f_{\rm{max}} \gg 1000$, this means
they cannot also have a high probability of having their origin in single collisions.
It would only be inferred that disks with $f_{\rm{obs}}/f_{\rm{max}} \ll 1000$ could
have their origin in single collisions, but since it is also possible that these disks
are the result of steady state collisional evolution there is no need to invoke
a single collision to explain their presence, which is why Table \ref{tab:hot2}
simply concluded that it is "not impossible" that the disks of
HD113766 and HD12039 are the product of single collisions.
What equation (\ref{eq:pff2}) does indicate, however, is that it is possible for single 
collisions to cause disks to spend some fraction of their time at a
luminosity enhanced above the nominal maximum value $f_{\rm{max}}$, and
that this occurs more readily for disks at smaller radii and around
higher mass stars.
However, whether single collisions really do achieve an observable increase in
luminosity depends on the size distribution of the collisional fragments, for
which it must be remembered that equation (\ref{eq:pff2}) used an unrealistically
optimistic estimate.

\subsection{Are parent planetesimals coincident with dust?}
\label{ss:coincident}

For similar reasons to those in \S \ref{ss:singlecoll} it is also possible to
show that the parent planetesimals of the dust are extremely unlikely to
originate in a planetesimal belt that is coincident with the dust.
The reason is that the mass remaining in such a belt would be insufficient to
replenish the dust for a length of time commensurate with the statistic that
2\% of stars show this phenomenon.
The observed dust luminosity, assuming this is comprised of
dust of size $D_{\rm{bl}}$ which has a lifetime of $t_{\rm{c}}(D_{\rm{bl}})$
(equation \ref{eq:tcdust}), implies a mass loss rate due to mutual collisions
between the dust grains of:
\begin{equation}
  dM_{\rm{loss}}/dt = 1700 f_{\rm{obs}}^2 r^{0.5} L_\star M_\star^{-0.5} (r/dr),
  \label{eq:mloss}
\end{equation}
in $M_\oplus$/Myr, and this is independent of the collisional evolution
model of \S \ref{s:model}.
However, due to the collisional evolution of a planetesimal belt's largest
members, there is a maximum mass that can remain in a belt at the same radius
as the dust at this age, and this is given in equation (\ref{eq:mmax1}).
This means that if the observed dust originates in an event which, for
whatever reason, is causing planetesimals in a belt at the same radius as
the dust to be converted into dust, then this can last a maximum time of
$t(f>f_{\rm{obs}})=M_{\rm{max}}/dM_{\rm{loss}}/dt$ before the planetesimal
belt is completely exhausted.
These figures are given in Table \ref{tab:hot3} which shows that the longest
the type of transient event observed could be sustained in these systems is
under 1 Myr, under the assumptions about the planetesimal belts employed in the
rest of the paper.

A maximum duration of 1 Myr is not sufficient to explain the statistic that
2\% of sun-like stars exhibit this phenomenon, since the median age of such
stars is 5 Gyr, indicating a typical duration (even if this occurs
in multiple, shorter, events) of around 100 Myr.
Clearly a reservoir of mass is required in excess of that which it
is possible to retain so close to the star.

\subsection{Constraints on parent planetesimal belt}
\label{ss:constraints}
If we assume that the observed mass of hot dust originates in planetesimals that
were initially in a belt at a radius $r_{\rm{out}}$ which has properties like those
assumed in the rest of the paper,
and a fractional luminosity of $f_{\rm{out}}$,
then there are two main constraints on that belt.
First, assuming that this belt has been collisionally evolving for the age of
the star, then this belt cannot have more mass (or luminosity) than the maximum
that could possibly remain due to collisional processing, i.e., $f_{\rm{out}} < f_{\rm{max}}$
(equation \ref{eq:fmax1}).
Second, it must have sufficient mass remaining to feed the observed mass loss rate
for long enough to reproduce the statistic that 2\% of stars exhibit this phenomenon
which implies a total duration of $>100$ Myr.
For a belt to have enough mass to feed the observed hot dust
luminosity of $f_{\rm{obs}}$ at a radius $r$ for a total time of $t_{\rm{hot}}$ in Myr
requires the belt to have a luminosity of:
\begin{equation}
  f_{\rm{out}} > 710 t_{\rm{hot}} f_{\rm{obs}}^2 r_{\rm{out}}^{-2} r^{0.5}
                 D_{\rm{c}}^{-0.5} L_\star^{0.5} (dr/r)^{-1},
  \label{eq:fout}
\end{equation}
or rather, this is the luminosity it must have had before it was depleted.

\begin{figure*}
  \centering
  \begin{tabular}{cc}
     \hspace{-0.15in} \includegraphics[width=3.2in]{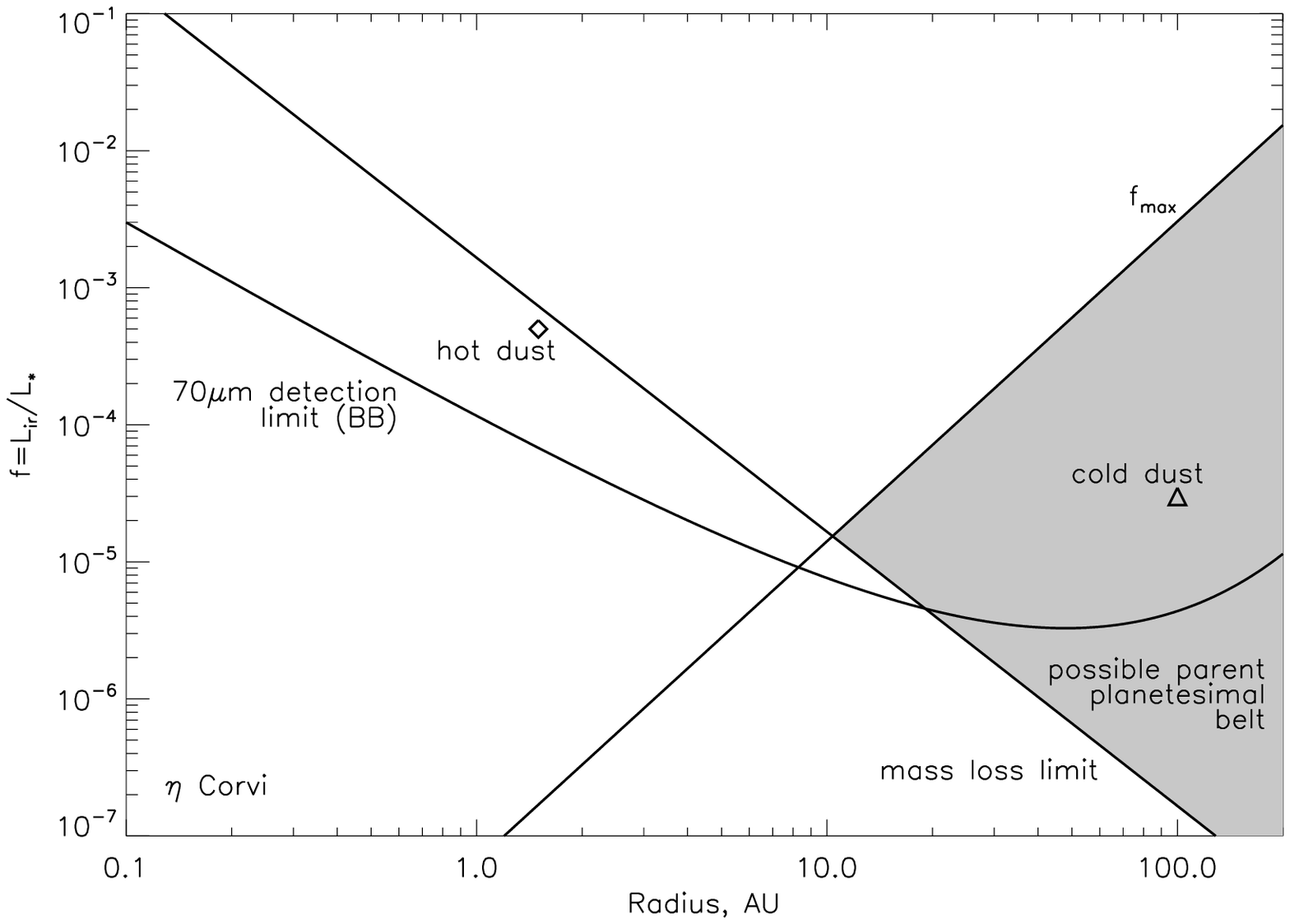} &
     \hspace{-0.15in} \includegraphics[width=3.2in]{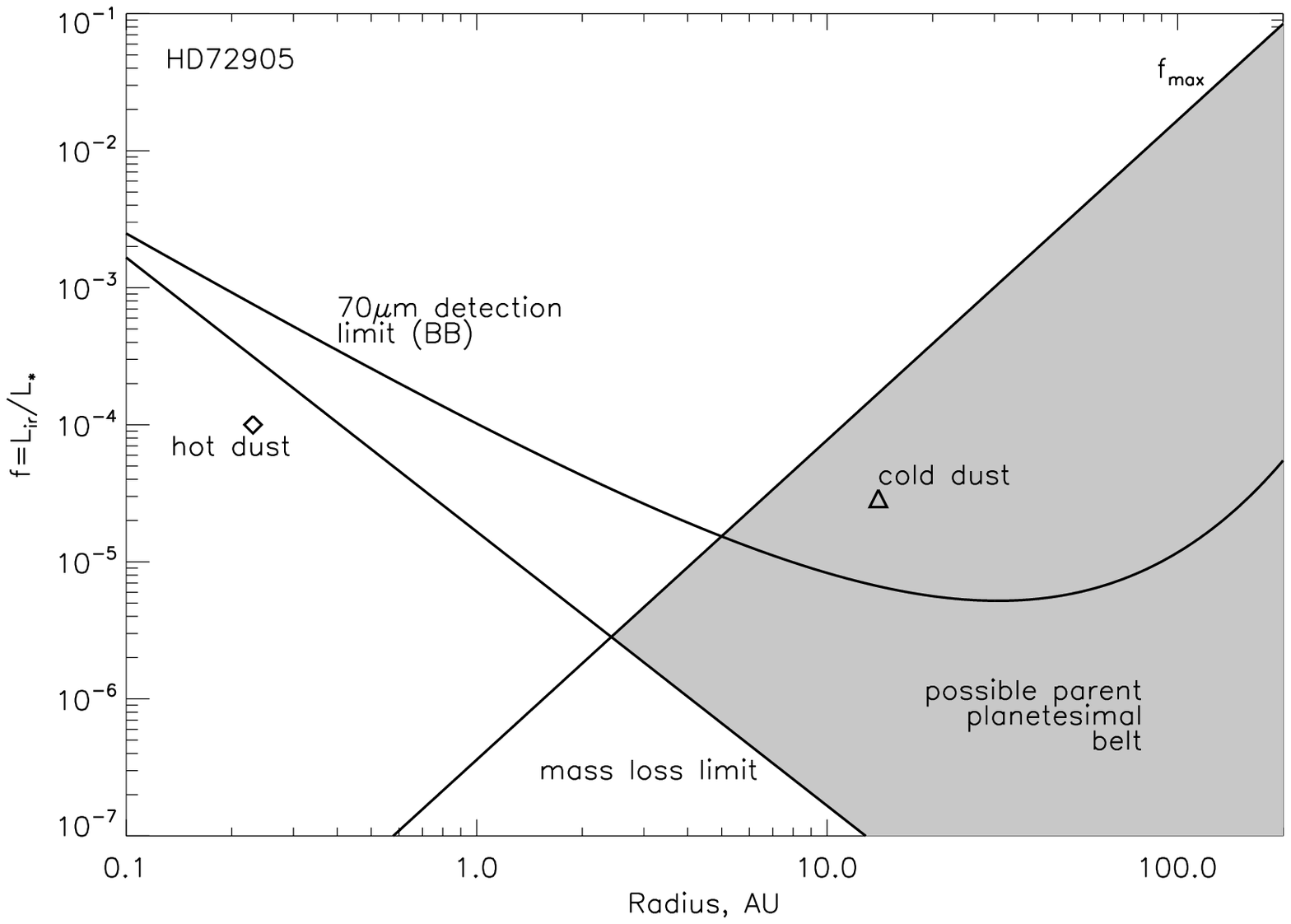} \\[-0.0in]
     \hspace{-0.15in} \includegraphics[width=3.2in]{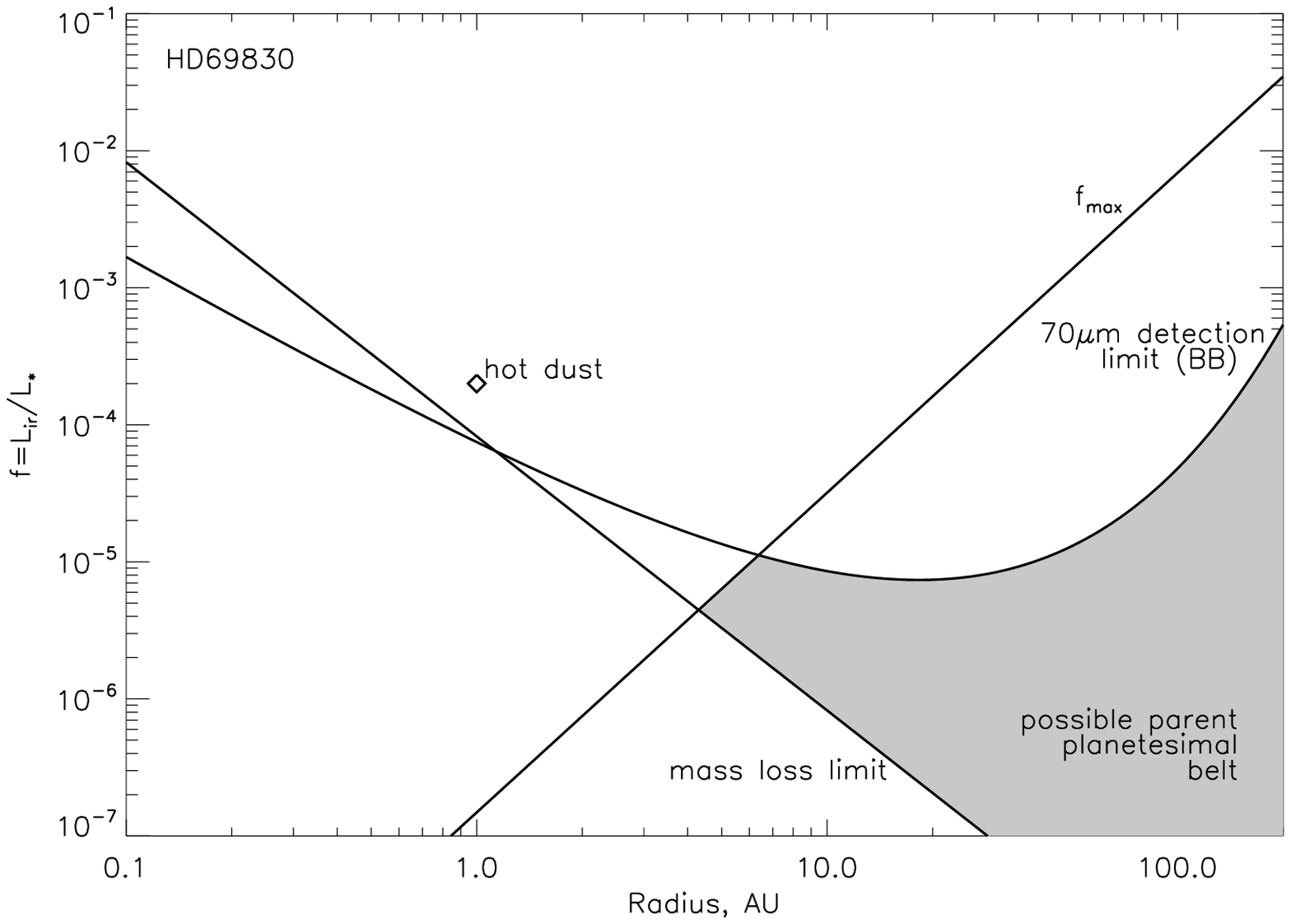} &
     \hspace{-0.15in} \includegraphics[width=3.2in]{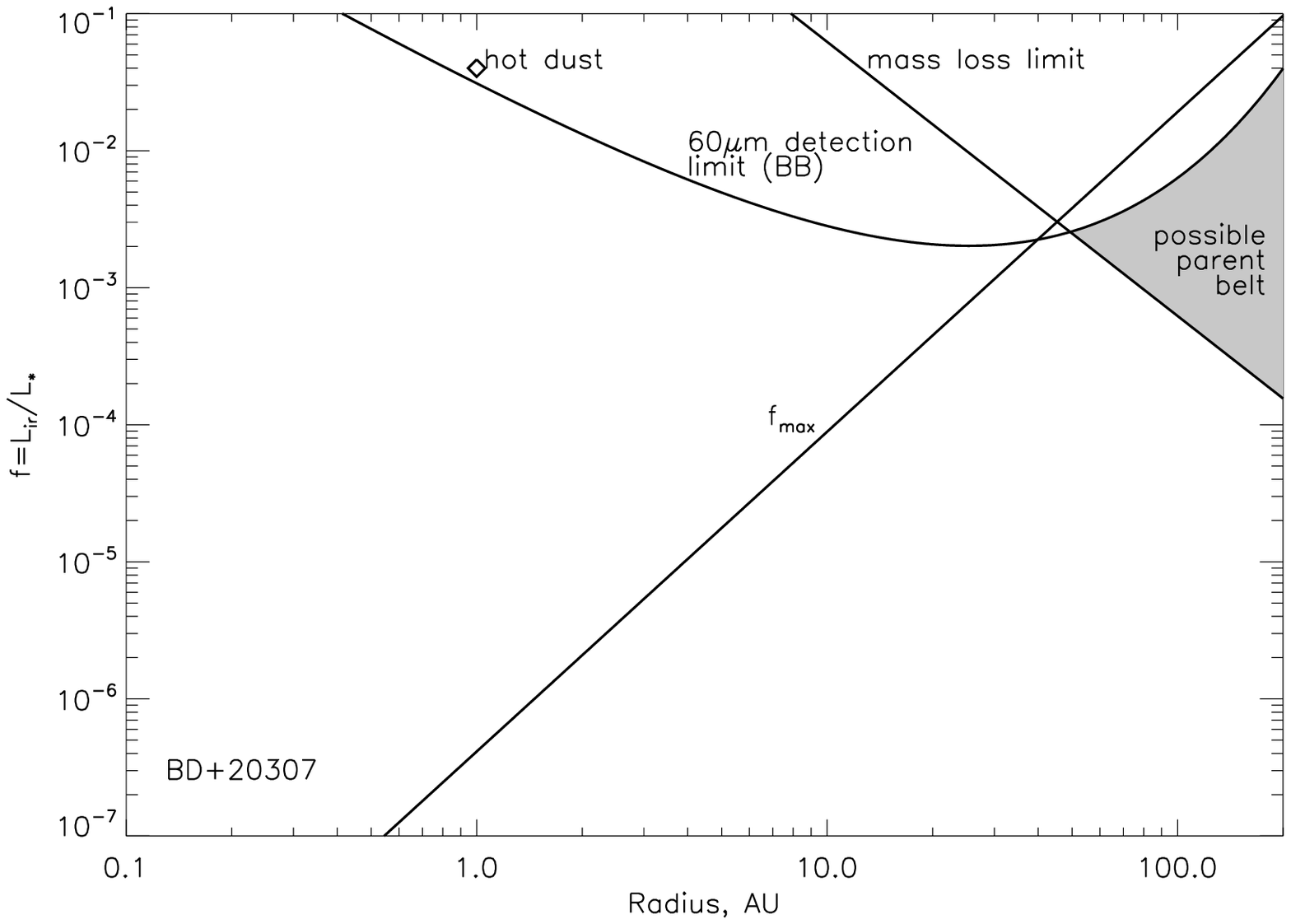} \\[-0.0in]
  \end{tabular}
  \caption{Constraints on the fractional luminosity and radius of the
  planetesimal belt feeding the observed transient hot dust (shaded region)
  for the systems: \textbf{(top left)} $\eta$ Corvi, \textbf{(top right)} HD72905,
  \textbf{(bottom left)} HD69830 and \textbf{(bottom right)} BD+20307.
  The solid lines are the constraints imposed by the far-IR detection limits (assuming
  black body emission), the maximum luminosity possible in a belt at this radius due to
  erosion by collisional processing, and the luminosity from a belt of sufficient mass
  to feed the observed mass loss rate for 100 Myr.
  The properties of the hot dust in these systems is shown with a diamond, and those
  of the cold dust, where known (the top two figures), shown with a triangle.}
  \label{fig:transsum}
\end{figure*}

Comparing this with the maximum mass possible at this age indicates that the parent
belt must have a minimum radius of:
\begin{eqnarray}
  r_{\rm{out}}(t_{\rm{hot}}) & > & 615t_{\rm{hot}}^{3/13}t_{\rm{age}}^{3/13}f_{\rm{obs}}^{6/13}
        r^{3/26} (dr/r)^{-6/13} \times  \nonumber \\
    & & D_{\rm{c}}^{-3/13}{Q_{\rm{D}}^\star}^{-5/26}e^{5/13}L_\star^{3/13}M_\star^{5/26}.
  \label{eq:rout}
\end{eqnarray}
Table \ref{tab:hot3} gives an estimate of the minimum radial location of such a
planetesimal belt, under the assumption that the event (or multiple events) of
high hot dust luminosity last $t_{\rm{hot}} = 100$ Myr.
These values indicate that the planetesimal belts must be at least a few AU from the star.
It must be cautioned that this conclusion is relatively weak in the case of HD69830,
since the uncertainty in the properties of the planetesimals still leaves two orders
of magnitude uncertainty in the maximum luminosity, $f_{\rm{max}}$,
and so also in the maximum mass $M_{\rm{max}}$ (see \S \ref{ss:pc}).
This means that, with suitable planetesimal belt properties, a belt in this
system that is coincident with the dust at 1 AU may be able to replenish the
observed phenomenon for 20 Myr.
However, we still consider this to be an unlikely scenario, since it would require
that the mass of the planetesimal belt is depleted at a constant rate for the
full 100 Myr, whereas most conceivable scenarios would result in a mass loss rate
which decreases with time as the planetesimal population is depleted thus requiring
an even larger starting mass.

These two constraints are summarized for the 4 systems with transient hot dust in
Fig.~\ref{fig:transsum}, which shows the shaded region of parameter space in
$f_{\rm{out}}$ and $r_{\rm{out}}$ where the parent planetesimal belt can lie.
This figure also shows the location of the hot dust at $f_{\rm{obs}}$ and $r$,
illustrating the conclusion of \S \ref{ss:qe} that this lies significantly
above $f_{\rm{max}}$, the maximum fractional luminosity expected for a planetesimal
belt at the age of the parent star.
Note that the value $r_{\rm{out}}(100\rm{Myr})$ given in Table \ref{tab:hot3}
denotes the intersection of the limits from $f_{\rm{max}}$ and from equation
(\ref{eq:fout}).

A third constraint for the parent planetesimal belt comes from far-IR observations
of these systems.
For 2/4 of the transient dust systems a colder dust component has already been detected:
$\eta$ Corvi has a planetesimal belt with a resolved radius of $\sim 100$ AU (Wyatt et al. 2005),
and HD72905 has one inferred to be at $\sim 14$ AU (Beichman et al. 2006a).
In both cases these outer planetesimal belts have been inferred to be at a
different spatial location from the hot dust either because of imaging constraints
(Wyatt et al. 2005) or from analysis of the SED (Beichman et al. 2006a).
The properties inferred for these planetesimal belts are indicated on Fig.~\ref{fig:transsum}
and lie within the shaded region, implying that these planetesimal belts
do not have to be transiently regenerated, and also provide a plausible source
population for the hot dust found closer in.
However, no such excess emission has been seen toward HD69830 at either 70 $\mu$m
(Beichman et al. 2005) or 850 $\mu$m (Sheret, Dent \& Wyatt 2004) indicating a 
planetesimal belt with a mass at most 5-50 times greater than our own Kuiper belt.
Likewise, BD+20307 does not have a detectable excess in IRAS 60 $\mu$m
observations (Song et al. 2005).

A low mass reservoir of planetesimals does not necessarily rule out the presence
of an outer planetesimal belt which is feeding the hot dust for two reasons.
First, the shaded region of Fig.~\ref{fig:transsum} actually constrains the properties of
the planetesimal belt at the time at which depletion started;
i.e., this population may have already been severely depleted by the same event
which is producing the dust and we are now nearing the end of the hot dust episode.
Second, the constraints imposed by a non-detection in the far-IR do not eliminate
the whole of the parameter space in which an outer planetesimal belt can lie.
Fig.~\ref{fig:transsum} includes the constraints on the outer planetesimal belt
imposed by the non-detection of excess in the far-IR, assuming that the
dust emits like a black body.
The resulting detection limit is then given by:
\begin{equation}
  f_{\rm{det}} = 3.4 \times 10^9 F_{\rm{det}}(\lambda)
                 d^2 r_{\rm{out}}^{-2}/B_\nu(\lambda,T_{\rm{bb}}), 
  \label{eq:fdet}
\end{equation}
where $F_{\rm{det}}$ is the detection limit in Jy, $d$ is the distance to the star in pc,
and $B_\nu(\lambda,T_{\rm{bb}})$ is in Jy/sr.
For BD+20307 the non-detection is limited by the sensitivity of
IRAS, and so lower limits should be achievable with Spitzer.
For the two systems with non-detections, the shaded region already takes
the far-IR constraint into account.

The simplification that the emission comes from black body-type grains
means that equation (\ref{eq:fdet}) underestimates the upper limit
from the far-IR fluxes.
This is because the majority of the luminosity comes from 
small grains which emit inefficiently at long wavelengths.
Indeed, the black body assumption would require the hot dust
of HD69830 and BD+20307 to have been detected in the far-IR,
whereas this is not the case.
We modeled the emission from non-porous silicate-organic refractory
grains in a collisional cascade size distribution at 1 AU from these
stars to find that the black body assumption used in equation (\ref{eq:fdet})
underestimates the limit by a factor of $3-5$ meaning that non-detection
of the hot dust in these systems in the far-IR is to be expected.
This also means that slightly more luminous outer planetesimal belts
than those indicated by the shaded region on Fig.~\ref{fig:transsum}
may still have escaped detection in the far-IR.

Until now we have not proposed a mechanism which converts the planetesimals
into dust.
Whereas Beichman et al. (2005) invoke sublimation of comets as the origin of
the hot dust, and use this to estimate the mass of the parent planetesimal
belt, we consider a scenario in which a significant fraction of material
of all sizes in the parent planetesimal belt is placed on orbits either
entirely coincident with the hot dust, or with pericenters at that location.
In this scenario the dust is reproduced in collisions and the material
maintains a collisional cascade size distribution.
Simply moving material from $r_{\rm{out}}$ to $r$ would result in an
increase in fractional luminosity from $f_{\rm{out}}$ to
$f_{\rm{out}}(r_{\rm{out}}/r)^2$.
This indicates that the parent planetesimal belt responsible for
the hot dust could have originally been on the line on Fig.~\ref{fig:transsum}
traced by $f_{\rm{out}} = f_{\rm{obs}} (r/r_{\rm{out}})^2$.
Since this is parallel to the mass loss limit line (equation \ref{eq:fout}),
and for all but HD69830 the observed hot dust component lies below this line,
this indicates that parent planetesimal belts in the shaded region
could be responsible for the hot dust observed, as long as a large fraction
of their mass is scattered in to the inner regions.
However, it is to be expected that only a fraction of the outer planetesimal
belt ends up in the hot dust region, and so it is more likely that the
parent planetesimal belt started on a line which falls off less steeply
than $\propto r_{\rm{out}}^{-2}$, and this is consistent with the ratio of the
hot and cold components of $\eta$ Corvi and HD72905 which indicate a
dependence of $f_{\rm{out}} = f_{\rm{obs}} (r/r_{\rm{out}})^{0.5 \pm 0.2}$;
it is also interesting to note that both have $r_{\rm{out}}/r =60-70$.
We defer further consideration of the expected properties of the
parent planetesimal belt to a more detailed model of the dynamics
of the types of events which could cause such a perturbation,
but simply note here that the existence of an outer planetesimal belt
is not ruled out by the current observational constraints in any
of the systems.

\section{Discussion}
\label{s:conc}
A simple model for the steady state evolution of dust luminosity for planetesimal
belts evolving due to collisions was described in \S \ref{s:model}.
This showed how at late times the remaining planetesimal belt mass and so
dust luminosity is independent of the initial mass of the belt.
This has important implications for the interpretation of the properties of detected disks.
This paper discussed the implications for the population of sun-like stars with hot dust
at $<10$ AU;
the implications for the statistics will be discussed in a forthcoming paper (Wyatt
et al., in prep.).

It was shown in \S \ref{ss:qe} that for 4/7 of the systems with hot dust their radius and age 
are incompatible with a planetesimal belt which has been evolving in quasi-steady state over the 
full age of the star, and in \S \ref{ss:pc} it was shown that this is the case
even when uncertainties in the model are taken into account.
This implies either that the cascade was started recently (within the last Myr or so), or that
the dust arises from some other transient event.
Recent ignition of the collisional cascade seems unlikely, since the mass required
to feed the observed luminosity would result in the growth of 2000 km planetesimals
which would stir the belt and ignite the cascade on timescales much shorter than
the age of the stars.
Possible origins for the transient event that have been proposed in the literature are:
recent collision between massive planetesimals in a planetesimal belt which introduces
dust with a size distribution $q \gg 11/6$ and so can be detected above a collisional
cascade which is too faint to detect;
one supercomet $\sim 2000$ km in diameter that was captured into a circular orbit
in the inner system replenishing the dust through sublimation (Beichman et al. 2005);
a swarm of comets scattered in from the outer reaches of the system (Beichman et al. 2005).
In \S \ref{ss:singlecoll} the collisional model was used to show that the transient
disks are very unlikely ($<0.001$\% for the most optimistic estimate for any of the
stars compared with a detection probability of 2\% for transient hot dust) to
have their origin in a recent collision;
such collisions occur too infrequently.
In \S \ref{ss:coincident} it was also shown that the parent planetesimals of the
observed dust must originate in a planetesimal belt much further from the star
than the observed dust, typically at $\gg 2$ AU.
This is because collisional processing means that the mass that can remain so
close to the star at late times is insufficient to feed the observed phenomenon.

The most likely scenario is thus a recent event which provoked one or more planetesimals to
be scattered in from further out in the disk (Beichman et al. 2005).
The observed dust could have been produced from such a scattered planetesimal population
through their grinding down in mutual collisions (\S \ref{ss:constraints}), although 
sublimation close to the pericenters of the planetesimals' orbits is a further possible source 
of dust.
More detailed study of the scattering and consequent dust production processes is
required to assess these possibilities.
However, this scenario is supported by the presence of far-IR emission originating from
a colder outer planetesimal belt component in 2/4 of the transient dust systems.
The constraints on the outer planetesimal belt which is feeding the
phenomenon are discussed in \S \ref{ss:constraints}, showing that
the outer planetesimal belts already found in $\eta$ Corvi and HD72905 provide a
plausible source population for the hot dust found closer in, and that the current
non-detection of cold dust around the remaining two systems does not rule out the
presence of an outer planetesimal belt capable of feeding the observed hot dust luminosity.

One clue to the origin of the parent planetesimals of the dust may be the composition
of that dust.
Silicate features have been detected in the mid-IR spectrum of all of the transient
hot dust stars (Song et al. 2005; Beichman et al. 2005; Beichman et al. 2006a;
Chen et al. 2006).
Detailed modeling of the spectrum of HD69830 indicates that the mineralogical
composition of its dust is substantially different from that of comets, rather
there is a close match to the composition of P- or D-type asteroids found mainly in
the 3-5 AU region of the solar system (Lisse et al. 2006).
While the radial location at which planetesimals of this composition form in the
HD69830 system will depend on the properties of its protostellar nebula, which may be 
significantly different to that of the protosolar nebula, as well as on the structure
and evolution of its planetary system, evidence for water ice in the dust spectrum
indicates that the parent body formed beyond the ice-line in this system
(Lisse et al. 2006), i.e., beyond $2-5.5$ AU (Lecar et al. 2006; Alibert et al. 2006).
Thus the compositional data supports the conclusion that the dust is not produced by a 
planetesimal that formed in situ.
However, it is worth noting that the same compositional data also
finds evidence for differentiation in the parent body (inferred from
abundance differences between the dust and the star) and for heating of
its rocky material to $>900$ K (inferred from the absence of amorphous
pyroxene), which would also have to be explained in the context of an outer
planetesimal belt origin for the dust.

An analogous transient event is thought to have happened in the solar system resulting in
the period known as the Late Heavy Bombardment (LHB) when the terrestrial planets
were subjected to an abnormally high impact rate from asteroids and comets.
This is believed to have been triggered by a dynamical instability in the
planetary system resulting from Jupiter and Saturn crossing the 1:2 resonance
during their slow migration (inwards for Jupiter, outwards for Saturn)
due to angular momentum exchange with the primordial Kuiper belt
(Gomes et al. 2005).
In this scenario both the asteroid and Kuiper belts were depleted with
a large fraction of these objects being scattered into the terrestrial planet
region during an event which lasted $10-150$ Myr (Gomes et al. 2005), i.e.,
exactly the type of event required to explain the observed hot dust in the scenario
proposed here (\S \ref{ss:constraints}).
Dynamical instabilities in extrasolar planetary systems can also arise from
mutual gravitational perturbations between giant planets which formed close
together (Lin \& Ida 1997; Thommes, Duncan \& Levison 1999).
In both scenarios slow diffusion of the orbits of the planets
means that the dynamical instability can occur up to several Gyr after
the formation of the planetary system.
The delay to the onset of the instability is determined by the separation
of the outer planet from the outer planetesimal belt (Gomes et al. 2005), or
from the separation between the planets (Lin \& Ida 1997), with larger separations
resulting in longer timescales.

Little is known about the planetary systems of four of the hot dust systems.
However, three Neptune mass (or Jupiter mass if the system is seen face-on)
planets have recently been discovered orbiting the star HD69830 at $<1$ AU
on nearly circular orbits (Lovis et al. 2006).
Dynamical simulations showed that the detected planetary system is stable on
timescales of 1 Gyr.
This does not, however, rule out the possibility of a dynamical instability having occurred.
While no mean motion or secular resonances are immediately identifiable within
the detected planetary system which could have have been crossed recently invoking
such a catastrophic event, it is possible that the instability arose with another
planet further out which has yet to be detected with longer timescale observations.
It is also possible that a fourth planet which existed in the region 0.19-0.63 AU
between the planets HD69830c and HD69830d has recently been scattered out due to a
dynamical instability (e.g., Thommes et al. 1999).
The region 0.3-0.5 AU was identified in Lovis et al. (2006) as being marginally stable,
and to encompass several mean motion resonances with the outer planet, including
the 1:2 resonance at 0.4 AU;
i.e., a putative fourth planet could have remained in this region for the past 2 Gyr
until the slow migration/diffusion of the outer planet (HD69830d) caused the 1:2
resonance to coincide with the orbit of the putative planet which was then scattered
outward thus promoting the depletion of an outer planetesimal belt much of which
was scattered into the inner regions of the system.
Alibert et al. (2006) considered that the most plausible formation scenario for the
planetary system of HD69830 included the inward migration of the outer planets
from beyond the ice-line at a few AU.
This would put a substantial distance between the outer planet (HD69830d) and any
outer planetesimal belt which favors a delay of 2 Gyr before the onset of the
instability.
Searches for further planetary companions in this system, and for the 
relic of its outer planetesimal belt, are clearly necessary to constrain the
evolutionary history of this system.

In conclusion, $\sim 2$\% of sun-like stars exhibit transient hot
dust in the terrestrial planet region;
this dust must originate in a planetesimal belt located further from
the star than the dust, typically at $\gg 2$ AU.
Just four members of this class are currently known, although it seems
reasonable to assume that our own solar system would have been placed
in this class during the LHB.
The frequency of this phenomenon indicates that either all stars are
subjected to an epoch of similar duration (lasting $\sim 100$ Myr assuming
a typical age of 5 Gyr) or that a smaller fraction of stars undergo much
longer (or multiple) events.
The distribution of the ages of the stars in this class indicate that the likelihood of
these events occurring falls off roughly inversely proportional to the
age of the stars.
An origin for these events in a dynamical instability as proposed for the LHB
in the solar system is supported by the recent discovery of a multiple
planet system coincident with the dust in one of the systems currently in
this class.
However, since the LHB in the solar system is thought to have lasted just
$\sim 100$ Myr, it remains to be seen whether we are to infer that dynamically
unstable planetary systems form around all stars, or that the LHB event in
other systems lasted much longer than in our own, or perhaps that there is in
fact more than one mechanism causing this hot dust signature.
Observations that further constrain the planet, planetesimal and dust
complements of the transient hot dust systems are needed to ascertain the 
similarities and dissimilarities within this population.

\acknowledgements
We are grateful for support provided by the Royal Society (MCW) and PPARC (RS).
We are also grateful to Ben Zuckerman, Joseph Rhee and Inseok Song for pointing
out that there is a strong (unrelated) infrared source in the vicinity of
HD128400 which is causing the excess identified by Gaidos (1999).

\appendix

\section{Summary of symbols}

The symbols which are employed in this paper are summarized in Table \ref{tab:symb}
along with the units assumed throughout the paper.

\begin{deluxetable*}{ccl}
  \tabletypesize{\scriptsize}
  \tablecaption{Symbols employed in this paper and their units.\label{tab:symb}}
  \tablewidth{0pt}
  \tablehead{ \colhead{Symbol}  & \colhead{Units}  & \colhead{Meaning} }
  \startdata
  $B_\nu(\lambda,T)$    & Jy/sr        & black body emission spectrum \\
  $d$                   & pc           & distance to star \\
  $dM_{\rm{loss}}/dt$   & $M_\oplus$/Myr & rate of mass loss assuming observed dust has size $D_{\rm{bl}}$ \\ 
  $dr$                  & AU           & planetesimal belt width \\
  $D_{\rm{bl}}$         & $\mu$m       & diameter of smallest dust in cascade \\
  $D_{\rm{c}}$          & km           & diameter of largest planetesimal in cascade \\
  $D_{\rm{cc}}$         & km           & smallest planetesimal capable of destroying planetesimals of diameter $D_{\rm{c}}$ \\
  $D_{\rm{pb}}$         & km           & minimum diameter of parent body required to produce observed dust \\
  $D_{\rm{t}}$          & km           & planetesimal diameter at transition between strength and gravity regimes \\
  $e$                   &              & mean orbital eccentricity of planetesimals \\
  $f$                   &              & fractional luminosity (= $L_{\rm{ir}}/L_\star$) in model \\
  $f_{\rm{det}}$        &              & fractional luminosity for emission from belt to be detected \\ 
  $f_{\rm{max}}$        &              & maximum fractional luminosity of cascade after time $t_{\rm{age}}$ \\
  $f_{\rm{obs}}$        &              & fractional luminosity observed \\
  $f_{\rm{out}}$        &              & fractional luminosity of putative outer planetesimal belt feeding the dust \\
  $f(e,I)$              &              & ratio of collision velocity to Keplerian velocity \\
  $f_{\rm{cc}}$         &              & see equation (\ref{eq:fcc}) \\
  $F_{\rm{det}}(\lambda)$& Jy           & detection limit at wavelength $\lambda$ \\
  $G(q,X_{\rm{c}})$     &              & see equation (\ref{eq:qgxc}) \\
  $I$                   & rad          & mean orbital inclination of planetesimals \\
  $K$                   &              & scaling factor in size distribution \\
  $L_\star$             & $L_\odot$    & stellar luminosity \\
  $L_{\rm{ir}}$         & $L_\odot$    & infrared luminosity of material in the cascade \\
  $M_\star$             & $M_\odot$    & stellar mass \\
  $M_{\rm{max}}$        & $M_\oplus$   & maximum mass remaining in cascade after time $t_{\rm{age}}$ \\
  $M_{\rm{tot}}$        & $M_\oplus$   & total mass of material in cascade \\
  $n(D)$                &              & size distribution of material in the cascade \\
  $n(D>D_{\rm{pb}})$    &              & number of objects in cascade larger than $D_{\rm{pb}}$ \\
  $dN_{\rm{c}}(D>D_{\rm{pb}})/dt$ &  Myr$^{-1}$ & destructive collision rate for planetesimals larger than $D_{\rm{pb}}$ \\
  $P(f>f_{\rm{obs}})$   &              & fraction of time collisions result in $f>f_{\rm{obs}}$ \\
  $q$                   &              & slope of size distribution \\
  $q_{\rm{g}}$          &              & slope of size distribution expected in the gravity regime \\
  $q_{\rm{s}}$          &              & slope of size distribution expected in the strength regime \\
  $Q_{\rm{D}}^\star$    & J kg$^{-1}$  & specific incident energy required to catastrophically destroy a planetesimal \\
  $r$                   & AU           & planetesimal belt radius, assumed to be coincident with dust \\
  $r_{\rm{out}}(t_{\rm{hot}})$ & AU    & outer planetesimal belt radius required to maintain $f_{\rm{obs}}$ for $t_{\rm{hot}}$ \\
  $r_{\rm{out}}$        & AU           & radius of putative outer planetesimal belt feeding the dust \\
  $s$                   &              & exponent in relation $Q_{\rm{D}}^\star \propto D^s$ \\
  $t(f>f_{\rm{obs}})$   & Myr          & time a planetesimal belt at $r$ can sustain $f>f_{\rm{obs}}$ \\
  $t_{\rm{age}}$        & Myr          & time since cascade initiated (assumed to be stellar age) \\
  $t_{\rm{c}}$          & yr           & collisional lifetime of planetesimals of size $D_{\rm{c}}$ \\
  $t_{\rm{c}}(D)$       & yr           & collisional lifetime of material of size $D$ \\
  $t_{\rm{hot}}$        & Myr          & total duration of hot episodes throughout stellar lifetime \\
  $v_{\rm{rel}}$        & m s$^{-1}$   & relative velocity of collisions \\
  $v_{\rm{esc}}$        & m s$^{-1}$   & escape velocity \\
  $v_{\rm{k}}$          & m s$^{-1}$   & Keplerian velocity \\
  $x_{\rm{t}}$          &              & jump in size distribution expected at $D_{\rm{t}}$ \\
  $X_{\rm{c}}$          &              & $=D_{\rm{cc}}/D_{\rm{c}}$ \\
  $\rho$                & kg m$^{-3}$  & planetesimal density \\
  $\sigma_{\rm{tot}}$   & AU$^2$       & total cross-sectional area of material in cascade
  \enddata
\end{deluxetable*}


\end{document}